%% file: paper.tex
\documentclass{article}
\usepackage{etex}

\usepackage{comment}
\usepackage{tikz}
\usetikzlibrary{matrix,arrows,decorations.pathmorphing,patterns,positioning}
\usepackage{amssymb}
\usepackage{ifthen}
\usepackage{color}
\usepackage{colortbl}
\usepackage{rotating}
\usepackage{multirow}
\usepackage{multicol}
\usepackage{url}

\usepackage{pgfplots}

\usepackage{wrapfig}
\usepackage{subfigure}
\usepackage{array}
\usepackage{framed}

\usepackage{epsfig}
\usepackage{amsmath}
\usepackage{amsfonts}
\usepackage{amssymb}

\usepackage{algorithmic}

\usepackage{algorithm}

\usepackage[normalem]{ulem}

\usepackage{ifthen}
\usepackage{boxedminipage}
\usepackage{fancyvrb}

\usepackage{soul}
\usepackage{graphicx}
\usepackage{graphics}
\usepackage{moreverb}
\usepackage{rotating}
\usepackage{enumerate}

\usepackage{tabularx,booktabs}

\usetikzlibrary{arrows}

\usepackage{xspace}

\newcommand{\gemm}{{\sc gemm}\xspace}
\newcommand{\dgemm}{{\sc dgemm}}

\definecolor{darkred}{rgb}{0.44,0,0}
\definecolor{darkgreen}{rgb}{0,0.44,0}
\definecolor{darkblue}{rgb}{0,0,0.44}
\definecolor{enrique}{rgb}{0,0,0}

\newcommand{\Jc}{{\cal J}_c}
\newcommand{\Pc}{{\cal P}_c}
\newcommand{\Ic}{{\cal I}_c}
\newcommand{\Jr}{{\cal J}_r}

\newcommand{\Ir}{{\cal I}_r}

\input{flatex}

\input{macros}

\date{
\today
}

\usepackage{listings}
\usepackage{xcolor}
\usepackage{fancyvrb}
\fvset{tabsize=2}
\definecolor{mygreen}{rgb}{0,0.6,0}
\pgfplotsset{every axis/.append style={
        scaled y ticks = false, 
        scaled x ticks = false, 
        y tick label style={/pgf/number format/.cd, fixed, fixed zerofill,
                            int detect,1000 sep={\;},precision=3},
        x tick label style={/pgf/number format/.cd, fixed, fixed zerofill,
                            int detect, 1000 sep={},precision=3}
    }
}

\newcommand{\fromto}[2]{{\color{black} #2}}

\title{Implementing Strassen's Algorithm with BLIS\\[0.2in]
	\large FLAME Working Note \# 79}

\author{
Jianyu Huang\footnote{%
Department of Computer Science,
The University of Texas at Austin,
Email: {\tt \{jianyu,tms,rvdg@cs.utexas.utexas.edu\}}.}
\and
Tyler M. Smith\footnotemark[1] \footnote{%
Institute for Computational Engineering and Sciences,
The University of Texas at Austin.
}
\and
Greg M. Henry\footnote{%
Intel Corporation,
2111 NE 25th Avenue, Bldg JF1-13, Hillsboro, OR 97124-5961.
Email: {\tt greg.henry@intel.com}.
}
\and
Robert A. van de Geijn\footnotemark[1] \footnotemark[2] 
}

\date{April 16, 2016}

\newcommand{\NoShow}[1]{}

\begin{document}

\maketitle

\begin{abstract}
\input 00abstract
\end{abstract}

\input body

\subsection*{Additional information}

Additional information regarding BLIS and related projects can be found at 
\begin{center}
{\tt http://shpc.ices.utexas.edu}
\end{center}

\subsection*{Acknowledgments}


This work was supported in part by the National Science
Foundation under Award No. ACI-1148125/1340293
and by Intel Corp.~through an Intel\textregistered\ 
Parallel Computing Center.
Access to the Maverick and Stampede supercomputers administered by TACC
is gratefully acknowledged.

\noindent
{\em Any opinions, findings, and conclusions or recommendations expressed in this material are those of the author(s) and do not necessarily reflect the views of the National Science Foundation. }

\bibliographystyle{IEEEtran}

\bibliography{biblio}



\end{document}

%% file: flatex.tex
\newcolumntype{I}{!{\vrule width 1.5pt}}
\newlength\savedwidth
\newcommand\whline{\noalign{\global\savedwidth\arrayrulewidth
                            \global\arrayrulewidth 1.5pt}%
           \hline
           \noalign{\global\arrayrulewidth\savedwidth}}

\newboolean{IsWide}
\setboolean{IsWide}{true}

\newcommand{\FlaPartition}[2]{
\ifthenelse{\boolean{IsWide}}{{\bf partition } \hspace{-1em} #1 \hspace{-1em} #2}
{{\bf partition } \+ \\ #1 \+ \\ #2 \- \-}
}

\newcommand{\FlaRepartition}[2]{
\ifthenelse{\boolean{IsWide}}{{\bf repartition } \hspace{-1em} #1 \hspace{-1em} #2}
{{\bf repartition } \+ \\ #1 \+ \\ #2 \- \-}
}

\newcommand{\FlaContinue}[1]{
\ifthenelse{\boolean{IsWide}}{{\bf continue with } #1
}
{{\bf continue with } \+ \\ #1 \-
}
}

\newcommand{\blocksize}{1}

\newcommand{\repartitionings}{
\begin{minipage}[t]{2in}
\ \\
\ \\
\ \\
\end{minipage}
}

\newcommand{\repartitionsizes}{ \hspace{ 1.25in} }

\newcommand{\WSrepartition}{
\begin{minipage}[t]{2in}
\ifthenelse{ \equal{\blocksize}{1} }{}
{%
\ifthenelse{ \equal{\blocksize}{2} }{~}
{\bf Determine block size $ \blocksize $} \\
}
{\bf Repartition}
\begin{tabbing}
in \= in \= \+ \kill
\repartitionings \+ \\
{\bf where } \hspace*{-2ex} \repartitionsizes 
\end{tabbing}
\end{minipage}
}

\newcommand{\WSrepartitionNarrow}{
\begin{minipage}[t]{2.15in}
\ifthenelse{ \equal{\blocksize}{1} }{}
{%
\ifthenelse{ \equal{\blocksize}{blank} }{~}
{\bf Determine block size $ \blocksize $} \\
}
{\bf Repartition}
\begin{tabbing}
i \= i \= \+ \kill
\repartitionings \+ \\
{\bf where } \hspace*{-2ex} \repartitionsizes 
\end{tabbing}
\end{minipage}
}

%% file: macros.tex
\newcommand{\GFLOPS}{GFLOPS}
\newcommand{\Strassen}{Strassen}

\newcommand{\ABCstrassen}{\textbf{ABC \mbox{Strassen}}}
\newcommand{\ABXstrassen}{\textbf{AB \mbox{Strassen}}}
\newcommand{\XXXstrassen}{\textbf{Naive \mbox{Strassen}}}

\newcommand{\figref}[1]{Figure~\ref{#1}}

\newcommand{\eref}[1]{Equation~\eqref{#1}}

\newcommand{\myalgcc}[1]{ {\footnotesize\ttfamily\textcolor{red}{//{#1}}}}

\newcommand{\strassen}{\mbox{\sc Strassen}}

\newcommand{\GOTO}{{\sc GotoBLAS}}
\newcommand{\BLIS}{{\sc BLIS}}

\definecolor{darkgreen}{rgb}{0,0.5,0}

\newcommand{\cblue}[1]{\textcolor{blue}{#1}}
\newcommand{\cgreen}[1]{\textcolor{darkgreen}{#1}}

\newcommand{\rankk}{rank-k}
\newcommand{\rankb}{rank-$b$}


%% file: 00abstract.tex
We dispel with ``street wisdom'' regarding the practical implementation of Strassen's algorithm for matrix-matrix multiplication (DGEMM).  Conventional wisdom: it is only practical for very large matrices.  Our implementation is practical for small matrices.  Conventional wisdom:  the matrices being multiplied should be relatively square.  Our implementation is practical for rank-k updates, where k is relatively small (a shape of importance for libraries like LAPACK).  Conventional wisdom: it inherently requires substantial workspace.  Our implementation requires no workspace beyond buffers already incorporated into conventional high-performance DGEMM implementations.  Conventional wisdom: a Strassen DGEMM interface must pass in workspace.  Our implementation requires no such workspace and can be plug-compatible with the standard DGEMM interface.  Conventional wisdom: it is hard to demonstrate speedup on multi-core architectures.  Our implementation demonstrates speedup over conventional DGEMM even on an Intel\textregistered\ Xeon Phi\texttrademark\ coprocessor\footnotemark[1]\ utilizing 240 threads. We show how a distributed memory matrix-matrix multiplication also benefits from these advances.

%% file: body.tex
\section{Introduction}
\label{sec:introduction}

\input 01intro \label{s:intro}

\section{Standard Matrix-matrix Multiplication} \label{s:gemm}

\input 02standard

\section{Strassen's Algorithm\fromto{ Reloaded}{}} \label{s:reload}

\input 03strassen

\section{Implementation and Analysis} \label{s:analysis}

\input 04analysis

\section{Performance Experiments} \label{s:perf}

\input 05performance

\section{Conclusion} \label{s:conclusion}

\input 06conclusion

\NoShow{
	\subsection*{Additional information}

\phantom{
	\begin{minipage}{\textwidth}
For additional information on BLIS visit
\begin{center}
\end{center}
\end{minipage}
}
Omitted for double blind review.
}

%% file: 01intro.tex
\footnotetext[1]{
Intel, Xeon, and Intel Xeon Phi are trademarks of Intel Corporation in the U.S. and/or other countries.
}

Strassen's algorithm (\strassen)~\cite{Strassen} for matrix-matrix multiplication (\gemm) has fascinated theoreticians and practitioners alike since it was first published,
in 1969.
That paper demonstrated that multiplication of $ n \times n $ matrices can be achieved 
in less than the $ O( n^{3}) $ arithmetic operations required by a conventional formulation.
It has led to many variants that improve upon this result~\cite{winograd,bini1979,sch1981,smi2013} as well as practical implementations~\cite{StrassenDouglas,Huss-Lederman:1996:ISA:369028.369096,D'alberto:2011:EPM:2049662.2049664,StrassenBenson}.  
The method can yield a shorter execution time than the best conventional algorithm with a modest degradation in numerical stability~\cite{HighamBook,DemmelStrassenStable2007,BallardStrassenStable15} by only incorporating a few levels of recursion.

From 30,000 feet the algorithm can be described as shifting computation with submatrices from multiplications to additions, 
reducing the $ O( n^3 ) $ term at the expense of adding $ O( n^2 ) $ complexity.  For current architectures, of greater consequence is the additional memory movements that are incurred
when the algorithm is implemented in terms of a  conventional \gemm\ provided by a high-performance implementation through the Basic Linear Algebra Subprograms (BLAS)~\cite{BLAS3} interface.
A secondary concern has been the extra workspace that is required.  This simultaneously limits the size of problem that can be computed and makes it so an implementation is not plug-compatible with the standard calling sequence supported by the BLAS.

An important recent advance in the high-performance implementation of \gemm\ is the BLAS-like Library Instantiation Software (\BLIS{} framework)~\cite{BLIS1}, a careful refactoring of the best-known approach to implementing conventional \gemm\ introduced by Goto~\cite{Goto:2008:AHP}.  Of importance to the present paper are the building blocks that \BLIS\ exposes, minor modifications of which support a new approach to implementating \strassen.
This approach \fromto{greatly reduces}{changes} data movement between memory layers and \fromto{hence mitigates}{can thus mitigate} the negative impact of the additional lower order terms incurred by \strassen.  These building blocks have similarly been exploited to improve upon the performance of, for example, the computation of the K-Nearest Neighbor~\cite{Chenhan:SC15}\NoShow{and Tensor Contraction~\cite{DevinSC16} problem}.
\fromto{}{The result is a family of \strassen\ implementations, members of which attain superior performance depending on the sizes of the matrices.}

 The resulting \fromto{formulation}{family} improves upon prior implementations of \strassen\ in a number of surprising ways:
\begin{itemize}
	\item It \fromto{outperforms}{can outperform} classical \gemm\ even for small square matrices.
	\item
	It \fromto{achieves}{can achieve} high performance for \rankk{} updates (\gemm\  with a small ``inner matrix size''), a case of \gemm\ frequently encountered in the implementation of libraries like LAPACK~\cite{LAPACK3}.
	\item
	It \fromto{does}{needs} not require additional workspace.
	\item
	It \fromto{is easy to parallelize for multi-core and many-core architectures.}{can incorporate directly the multi-threading in traditional \gemm\ implementations.}
	\item
	It can be plug-compatible with the standard \gemm\ interface supported by the BLAS.
	\item 
	It can be incorporated into practical distributed memory implementations of \gemm.
\end{itemize}
Most of these advances run counter to conventional wisdom \fromto{.}{and are backed up by theoretical analysis and practical implementation.}

%% file: 02standard.tex
We start by discussing naive computation of matrix-matrix multiplication (\gemm), how it is supported as a library routine by the Basic Linear Algebra Subprograms (BLAS)~\cite{BLAS3}, 
\fromto{how blocking can be employed,}{} how modern implementations block for caches, and how that implementation supports multi-threaded parallelization.

\subsection{Computing $ C = \alpha A B + C $}

Consider $ C = \alpha A B + C $, where $ C $, $ A $,
and $ B $ are $ m \times n $, $ m \times k $, and $ k \times n $
matrices, respectively, and $\alpha$ is a scalar. If 
the $ (i,j) $ entry of $ C $, $ A $, and $ B $ are respectively denoted by
$ \gamma_{i,j} $, $ \alpha_{i,j} $, and $ \beta_{i,j} $,
\NoShow{
{\setlength{\arraycolsep}{2pt}
\[
C = \left( \begin{array}{c c c}
\gamma_{0,0} & \cdots & \gamma_{0,n-1} \\
\vdots  & & \vdots \\
\gamma_{m-1,0} & \cdots & \gamma_{m-1,n-1} 
\end{array} \right),
A = \left( \begin{array}{c c c}
\alpha_{0,0} & \cdots & \alpha_{0,k-1} \\
\vdots  & & \vdots \\
\alpha_{m-1,0} & \cdots & \alpha_{m-1,k-1} 
\end{array} \right), 
\mbox{and }
B = \left( \begin{array}{c c c}
\beta_{0,0} & \cdots & \beta_{0,n-1} \\
\vdots  & & \vdots \\
\beta_{k-1,0} & \cdots & \beta_{m-1,k-1} 
\end{array} \right),
\]%
}
}
then computing $ C= \alpha A B + C $ is achieved by 
\[
 \gamma_{i,j} = \alpha \sum_{p=0}^{k-1} \alpha_{i,p} \beta_{p,j} + \gamma_{i,j},
 \]
which requires 
$ 2 m  n  k $ floating point operations (flops).

\subsection{Level-3 BLAS matrix-matrix multiplication}

(General) matrix-matrix multiplication (\gemm) is supported in the level-3 BLAS~\cite{BLAS3} interface as
\begin{quote}
	\footnotesize
	\tt DGEMM( \begin{tabular}[t]{@{}l}
			transa, transb, m, n, k, alpha, \\
	\tt ~~~~~~~A, lda, B, ldb, beta, C, ldc )
	\end{tabular}
\end{quote}
where we focus on double precision arithmetic and data. 
\fromto{It}{This call} supports 
\[
\begin{array}{l l}
C = \alpha A B + \beta C, &
C = \alpha A^T B + \beta C, \\
C = \alpha A B^T + \beta C, &\mbox{and }
C = \alpha A^T B^T + \beta C 
\end{array}
\] 
depending on the choice of {\tt transa} and {\tt transb}. 
In our discussion we can assume $ \beta = 1 $ since $ C $ can always first be multiplied by that scalar as a preprocessing step, which requires only $ O( n^2 ) $ flops.  
Also, by internally allowing both a row stride and a column stride for {\tt A}, {\tt B}, and {\tt C} (as the \BLIS{} framework does), transposition can be easily supported by swapping these strides. 
It suffices then to consider
$
C = \alpha A B + C 
$. 
 
\subsection{Computing with submatrices}

Important to our discussion is that
we partition the matrices and stage the matrix-multiplication as computations with submatrices.
For example, 
let us assume that $ m $, $ n $, and $ k $ are all even and partition %
{\setlength{\arraycolsep}{2pt} 
\[ 
C = \left( \begin{array}{c c} 
C_{00} & C_{01} \\ 
C_{10} & C_{11} 
\end{array}
\right)  
, 
A = \left( \begin{array}{c c} 
A_{00} & A_{01} \\ 
A_{10} & A_{11} 
\end{array}
\right)  
, 
B = \left( \begin{array}{c c} 
B_{00} & B_{01} \\ 
B_{10} & B_{11} 
\end{array}
\right),
\] %
}%
where $ C_{00} $ is $ \frac{m}{2} \times \frac{n}{2} $, $ A_{00} $ 
is $ \frac{m}{2} \times \frac{k}{2} $, and $ B_{00} $  $ \frac{k}{2} \times \frac{n}{2} $. 
Then 
\[
\begin{array}{l c l}
C_{00} &=& \alpha (A_{00} B_{00} + A_{01} B_{10}) + C_{00} \\
C_{01} &=& \alpha (A_{00} B_{01} + A_{01} B_{11}) + C_{01} \\
C_{10} &=& \alpha (A_{10} B_{00} + A_{11} B_{10}) + C_{10} \\
C_{11} &=& \alpha (A_{10} B_{01} + A_{11} B_{11}) + C_{11} 
\end{array}
\]
computes $ C = \alpha A B + C $\fromto{,}{}
via eight multiplications and eight additions with submatrices, still requiring \fromto{}{approximately} $ 2 m n k $ flops.

\subsection{The GotoBLAS algorithm for \gemm}

\input fig_side_by_side

\figref{fig:side_by_side}(left) illustrates the way the \GOTO{} \cite{GOTOBLASweb} (predecessor of OpenBLAS \cite{OpenBLASweb}) approach structures the blocking for three layers of cache (L1, L2, and L3) \fromto{for}{when} computing $ C = A B + C $\fromto{}{,} as implemented in \BLIS.
For details we suggest the reader consult the papers on the \GOTO{} \gemm~\cite{Goto:2008:AHP} and \BLIS~\cite{BLIS1}. 
\fromto{We will revisit the modified version of \GOTO{} for \strassen\ in Section \ref{s:stra}.}{In that figure, the indicated block sizes $m_C$, $ n_C $, and $ k_C $ are chosen so that submatrices fit in the various caches while $m_R$ and $ n_R $ relate to the size of contributions to $ C $ that fits in registers.  For details on how these are chosen, see~\cite{BLIS1,BLIS5}.}


Importantly,  
\begin{itemize}
	\item 
The row panels $ B_p $ that fit in the L3 cache%
\footnote{If an architecture does not have an L3 cache, this panel is still packed to make the data contiguous and to reduce the number of TLB entries used.}  
are packed into contiguous memory, yielding $ \widetilde B_p $.  
\item 
Blocks $ A_i $ that fit in the L2 cache are packed into buffer $ \widetilde A_i $. 
\end{itemize} 
It is \fromto{}{in part} this \emph{packing} that we are going to exploit as we implement one or more levels of \strassen. 

\subsection{Multi-threaded implementation}

\fromto{The \GOTO{} and \BLIS{} implementations of \gemm\ differ in that} \BLIS{} exposes all the illustrated loops, requiring only the micro-kernel to be optimized for a given architecture\fromto{, while }{. In contrast,} in the \GOTO\ implementation the micro-kernel and the first two loops around it form an inner-kernel that is implemented as \fromto{}{a} unit. 
\fromto{One benefit is that}{As a result,} the \BLIS\ implementation exposes five loops (two more than the \GOTO\ implementation) that can be parallelized, as discussed in~\cite{BLIS3}.  In this work, we mimic the insights from that paper.

%% file: fig_side_by_side.tex
\begin{figure*}[tb!]
~
\vspace{-0.5in}
\begin{center}
\includegraphics[width=1.06\textwidth]{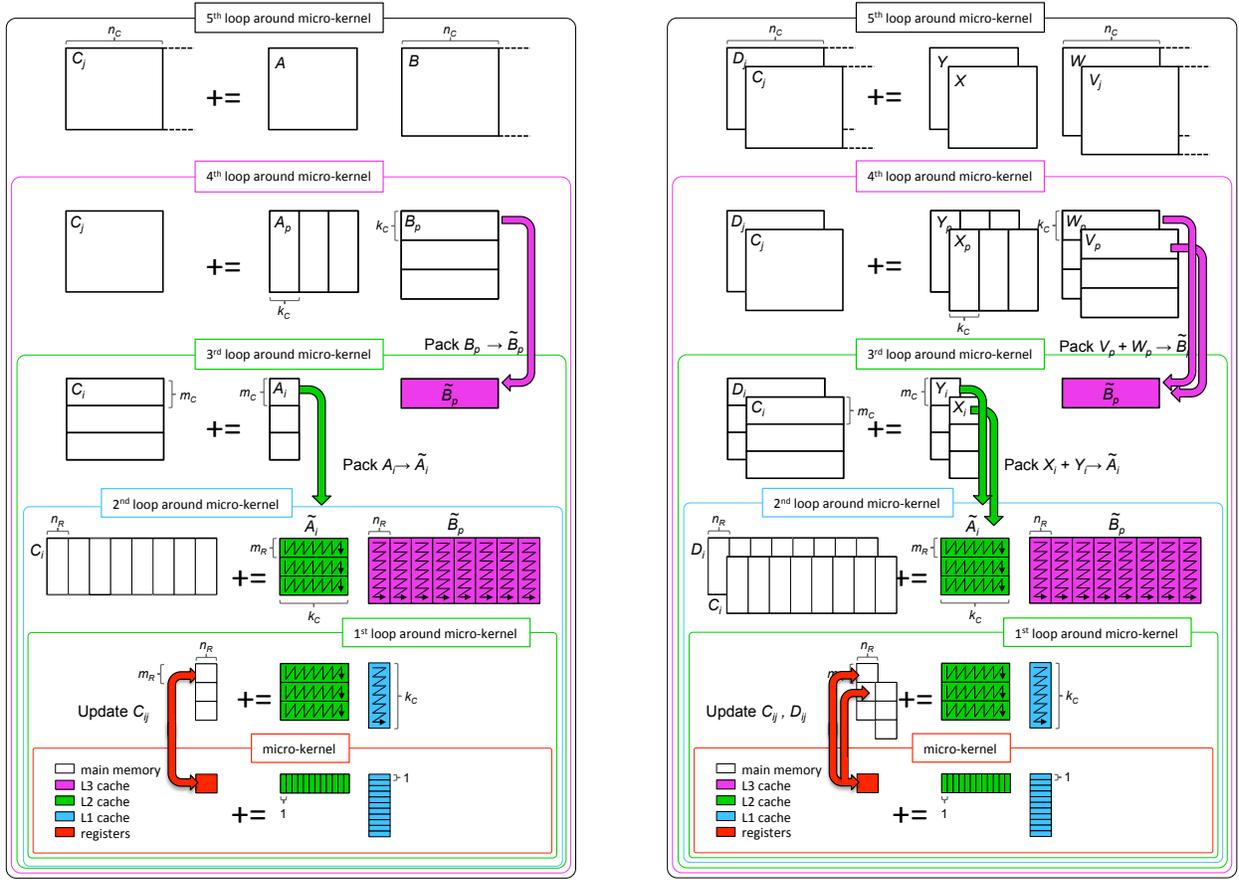}
\end{center}
\vspace{-0.55in}
\caption{
	Left: Illustration (adapted from~\cite{BLIS5} with permission of the authors) of the BLIS implementation of the \GOTO{} \gemm\  algorithm.  All computation is cast in terms of a micro-kernel that is highly optimized.    Right: modification that implements the representative computation $ M = ( X + Y)( V+W); C +\!\!= M; D +\!\!= M $ of general operation \eqref{e:straprim1}.
}
\label{fig:side_by_side}
\end{figure*}

%% file: 03strassen.tex
In this section, we present the basic idea and practical considerations of \strassen%
\fromto{. We reload \Strassen's algorithm  and decompose any level of \strassen}{, decomposing it} into a combination of general operations\fromto{, which}{\ that} can be adapted to
the high-performance implementation of a traditional \fromto{matrix matrix multiplication}{\gemm}. 

\subsection{The basic idea}
 
\begin{figure}[tb!]
{
	\footnotesize%
\begin{equation*}
\begin{array}{l @{\hspace{1pt}} c @{\hspace{1pt}} l l r}  
M_0 &=& ( A_{00} + A_{11} ) ( B_{00} + B_{11} ); 
& 
C_{00} +\!\!= \alpha M_0;  C_{11} +\!\!= \alpha M_0;  \\   
M_1 &=& ( A_{10} + A_{11} ) B_{00}; 
& 
C_{10} +\!\!= \alpha M_1 ;  C_{11} -\!\!= \alpha M_1 ; \\  
M_2 &=&  A_{00} ( B_{01} - B_{11} ); 
&  
C_{01} +\!\!= \alpha M_2 ;  C_{11} +\!\!= \alpha M_2 ;
\\  
M_3 &=& A_{11}( B_{10} - B_{00} ); 
&  
C_{00} +\!\!=  \alpha M_3 ;  C_{10} +\!\!= \alpha M_3 ;
\\  
M_4 &=& ( A_{00} + A_{01}) B_{11};  
& 
C_{01} +\!\!=  \alpha M_4 ;   C_{00} -\!\!= \alpha M_4;
\\  
M_5&=& (A_{10} - A_{00} )( B_{00} + B_{01} ); 
& 
C_{11} +\!\!= \alpha M_5;  
\\  
M_6&=& (A_{01} - A_{11} )( B_{10} + B_{11} );  
& 
C_{00} +\!\!= \alpha M_6 ; 
\end{array}
\label{eqn:allops}
\end{equation*} 
} 
\caption{All operations for one-level \strassen{}. Note that each row is a special case of general operation (\ref{e:straprim1}).}
\label{fig:strassenlevel1}
\end{figure}

It can be  
verified that the operations in \figref{fig:strassenlevel1}
also compute $ C = \alpha A B + C $, requiring only seven multiplications with submatrices.
The computational cost is, approximately, reduced from $ 2 mnk $ flops 
to $ ({7}/{8}) 2mnk $ flops, at the expense of a lower order number of extra
additions.  
\figref{fig:strassenlevel1}  describes \fromto{one level of \Strassen{}'s algorithm, which}{what} we will call {\em one-level \strassen}.

\subsection{Classic \Strassen's algorithm}

Each of the matrix multiplications that computes an intermediate
result $ M_k $ can itself be computed with another level of \Strassen{}'s
algorithm.  This can then be repeated recursively.

If originally $ m = n = k = 2^d $, where $ d $ is an integer, then the cost becomes
\[
 \left( {7}/{8} \right)^{\log_2(n)} 2 n^3 =
  n^{\log_2(7/8)}  2 n^3 \approx
2 n^{2.807} \mbox{~flops}.
\]
In this discussion, we ignored the increase in the total
number of extra additions, which turns out to contribute a lower order term.

\subsection{Practical considerations}



A high-performance implementation of a traditional matrix-matrix
multiplication requires careful attention to details related to data
movements between memory layers, scheduling of operations, and
implementations at a very low level (often in assembly code).
Practical implementations \fromto{of \Strassen{}'s
algorithm}{recursively} perform a few levels of \strassen\ until the matrices
become small enough so that a traditional high-performance
\dgemm\ is faster\fromto{ than continuing the recursion}{}.  At that point, the recursion stops and a high-performance \dgemm\ is used for the subproblems.
\fromto{However,}{In prior implementations,} the switch point is \fromto{}{usually} as large as 2000 for double precision square matrices on a single core of an x86 CPU\fromto{}{~\cite{D'alberto:2011:EPM:2049662.2049664,StrassenBenson}}. \fromto{We show that the}{We will see that, for the same architecture, one of our implementations has a} switch point\fromto{can be}{} as small as 500
(\figref{fig:model}).
\fromto{ in Section \ref{s:perf}}{}

In an ordinary matrix-matrix multiplication, three matrices must be
stored, for a total of $ 3 n^2 $ floating point numbers (assuming all matrices are $ n \times n $).
The most naive implementation of one-level \strassen\ requires an additional
seven submatrices of size $ \frac{n}{2} \times \frac{n}{2} $ (for $ M_0 $ through $ M_6
$)  and ten matrices of size $ \frac{n}{2} \times \frac{n}{2} $ for $ A_{00} +
A_{11} $, $ B_{00} + B_{11} $, etc.
A careful ordering of the computation can reduce this to
\fromto{2}{two} matrices~\cite{BoyerStrassen}. 
\fromto{However, that approach imposes too many dependencies\fromto{,}{\ and is} thus not practical for parallelization.}
We show that the computation can be organized so
  that no temporary storage beyond that required for a
  high-performance traditional \fromto{matrix-matrix multiplication}{\dgemm} is needed%
\fromto{ at the lowest level of the recursion}{}.
\fromto{Besides that}{In addition}, it is easy to \fromto{parallize}{parallelize} for multi-core and many-core architectures with our approach, since we can adopt the same parallel scheme \fromto{embeded in}{advocated by} \fromto{traditional matrix-matrix multiplication}{BLIS}.

The general case where one of more or the dimensions is not a convenient multiple of a power of two leads to the need to either pad matrices or to treat a remaining ``fringe'' carefully\fromto{}{~\cite{Huss-Lederman:1996:ISA:369028.369096}}. 
\NoShow{The implementation of GEMM in \BLIS\  also has to deal with \fromto{this issue}{fringes} when matrices are not nice multiples of the inherent block sizes encountered in the micro-kernel.  
For this reason, \fromto{}{in our implementations,} dealing with the fringe just involves index re-calculations in the packing buffer or micro-kernel for a high-performance traditional matrix multiplication, \fromto{with no extra padding buffer required}{requiring no extra padding}. 
}
Traditionally, it is necessary to pad $ m $, $ n $, and $ k $ to be integer multiples of two.  In our approach this can be handled internally by padding $ \widetilde A_i $ and $ \widetilde B_p $, and by using tiny ($ m_R \times n_R$) buffers for $ C $ along the fringes (much like the BLIS framework does).

\subsection{\fromto{General operations for \strassen}{One-level \strassen\  reloaded}}
\label{sec:hp:1level}

The operations summarized in \figref{fig:strassenlevel1} are all special cases of 
\begin{equation}
M = \alpha ( X + \delta Y)( V + \epsilon W ); ~~~C+\!\!= \gamma_0 M; ~~~D+\!\!= \gamma_1 M;
\label{e:straprim1}
\end{equation}
for appropriately \fromto{picked}{chosen} $ \gamma_0, \gamma_1, \delta, \epsilon \in \{ -1, 0, 1 \} $.  
\fromto{Note that}{Here,} $X$ and $Y$ are submatrices of $A$, $V$ and $W$ are submatrices of $B$, \fromto{}{and} $C$ and $D$ are submatrices of original $C$.
\NoShow{We will see next is that by incorporating the addition of matrices in the packing and updating $ C $ and $ D $ simultaneously, essentially no temporary storage beyond the buffers already incorporated in the standard \BLIS{} approach for \gemm\ is needed.  This eliminates the explicit addition of submatrices and the data movement between memory layers that is associated with such additions.
}
 
Let us focus on how to modify the algorithm illustrated in \figref{fig:side_by_side}(left) \fromto{that implements $ C = A B + C $}{} in order to accommodate the representative computation 
\[
 M = ( X + Y )( V + W ); C +\!\!= M; D +\!\!= M. 
\label{e:straprim2}
\]
As illustrated in \figref{fig:side_by_side}(right), the key insight is that the additions of matrices $ V + W $ can be incorporated in the packing into buffer $ \widetilde B_p $ and the additions of matrices $ X + Y $
in the packing into buffer $ \widetilde A_i $. 
Also\fromto{}{,} when a small block of $ ( X + Y )( V + W )$ is accumulated in registers\fromto{,}{} it can be added to the appropriate parts of both $ C $ and $ D $\fromto{.}{, multiplied by $ \alpha \gamma_0 $ and $ \alpha \gamma_1 $, as needed, inside a modified micro-kernel.}
This avoids multiple passes over the various matrices, which would otherwise add a considerable overhead \fromto{}{from memory movements}. 
\NoShow{\fromto{This is}{These insights are} illustrated in \figref{fig:side_by_side}(right). 
}
\fromto{In addition, we modify the packing part to further incorporate $ \delta $ and $ \epsilon $, and we update $ C $ and $ D $ matrices by $ \alpha \gamma_0 $ and $ \alpha \gamma_1 $ separately inside the micro-kernel.}{}

\subsection{Two-level \strassen\ \fromto{}{reloaded}}
\label{sec:hp:2level}

\input twolevels

Let %
{
	\setlength{\arraycolsep}{2pt}
\[ 
\begin{array}{c}
C  = \left( \begin{array}{c c | c c} 
C_{0,0} & C_{0,1} & C_{0,2} &C_{0,3} \\
C_{1,0} & C_{1,1} & C_{1,2} &C_{1,3} \\ \hline 
C_{2,0} & C_{2,1} & C_{2,2} &C_{2,3} \\
C_{3,0} & C_{3,1} & C_{3,2} &C_{3,3} \\
\end{array}
\right)  
, 
A = \left( \begin{array}{c c | c c}  
A_{0,0} & A_{0,1} & A_{0,2} &A_{0,3} \\ 
A_{1,0} & A_{1,1} & A_{1,2} &A_{1,3} \\ \hline  
A_{2,0} & A_{2,1} & A_{2,2} &A_{2,3} \\ 
A_{3,0} & A_{3,1} & A_{3,2} &A_{3,3} \\
\end{array}
\right),  
\mbox{ and }
B = \left( \begin{array}{c c | c c}   
B_{0,0} & B_{0,1} & B_{0,2} &B_{0,3} \\  
B_{1,0} & B_{1,1} & B_{1,2} &B_{1,3} \\ \hline   
B_{2,0} & B_{2,1} & B_{2,2} &B_{2,3} \\  
B_{3,0} & B_{3,1} & B_{3,2} &B_{3,3} \\
\end{array}
\right), 
\end{array} 
 \]%
}%
 where $ C_{i,j} $ is $ \frac{m}{4} \times \frac{n}{4} $, $ A_{i,p} $ 
is $ \frac{m}{4} \times \frac{k}{4} $, and $ B_{p,j} $ is $ \frac{k}{4} \times \frac{n}{4} $.  Then it can be  
verified that the computations in \figref{fig:2stepB}
compute $ C = \alpha A B + C $. 
The operations found there can be cast as special cases of 
\[
\begin{array}{rcl}
M = \alpha ( X_0 + \delta_1 X_1 + \delta_2 X_2 + \delta_3 X_3) \times \\
~~~~~~~( V_0 +
\epsilon_1 V_1 + \epsilon_2 V_2 + \epsilon_3 V_3 ); \\
C_0 +\!\!= \gamma_0 
M; C_1 +\!\!= \gamma_1 M; C_2 +\!\!= \gamma_2 M; C_3 +\!\!= \gamma_3 M 
\end{array}
\label{e:straprim3}
\]
by appropriately picking $ \gamma_i, \delta_i, \epsilon_i \in \{ -1, 0, 1 \} $. 
Importantly, the computation now requires 49 multiplications for submatrices as opposed to 64 for a conventional \gemm. 

To extend the insights from Section~\ref{sec:hp:1level} so as to integrate two-level \strassen{} into the \BLIS{} \gemm\ implementation, we incorporate the addition of up to four
submatrices of $A$ and $B$, the updates of up to four submatrices of $C$ inside the micro-kernel, and the tracking of up to four submatrices in the loops in \BLIS{}.

\subsection{Additional levels\fromto{ of \strassen}{}}

\fromto{From Sections~\ref{sec:hp:1level} and~\ref{sec:hp:2level} a}{A} pattern 
now emerges.  The operation needed to integrate $ k $ levels of \strassen{}
is given by 
\begin{equation}
\begin{array}{l}
M = \alpha \left( \sum_{s=0}^{l_X-1}
\delta_s X_s \right) 
\left( \sum_{t=0}^{l_V-1} \epsilon_t V_t
\right);  ~~~~~~~
C_r +\!\!= \gamma_r
M ~ \mbox{ for } r = 0 , \ldots, l_C-1. 
\end{array}
\label{e:straprim4}
\end{equation}
\NoShow{
A routine implementing this would want to order the terms so that 
\begin{eqnarray*}
\delta_i &=& \left\{ 
\begin{array}{ll}
1 \mbox{ or } -1 & \mbox{if } i < l_X \\
 0 &\mbox{if } i \geq l_X \\
\end{array}
\right.  \\
\epsilon_j &=& \left\{ 
\begin{array}{ll}
1 \mbox{ or } -1 & \mbox{if } j < l_V \\
 0 &\mbox{if } j \geq l_V \\
\end{array}
\right. 
\\
\gamma_p &=& \left\{ 
\begin{array}{ll}
1 \mbox{ or } -1 & \mbox{if } p < l_C \\
 0 &\mbox{if } p \geq l_C \\
\end{array}
\right. 
\end{eqnarray*}
}%
For each number, $ l $, of levels of \strassen{} that are integrated, a table can then be created that 
captures all the computations to be executed.

%% file: twolevels.tex
\begin{figure*}[tb!]
{
	\footnotesize
\[
\begin{array}{| l | l |} \hline
\begin{array}{l l l l l}
M_{0} = ( A_{0, 0} \!+\! A_{2, 2} \!+\! A_{1, 1} \!+\! A_{3, 3}  )  ( B_{0, 0} +
B_{2, 2} \!+\! B_{1, 1} \!+\! B_{3, 3}  ); &
 C_{0, 0} +\!\!= M_{0};
&
C_{1, 1} +\!\!= M_{0};
&
C_{2, 2} +\!\!= M_{0};
& 
C_{3, 3} +\!\!= M_{0};
\\
M_{1}  =  ( A_{1, 0} \!+\! A_{3, 2} \!+\! A_{1, 1} \!+\! A_{3, 3}  )  ( B_{0, 0} +
B_{2, 2}  );
&
  
C_{1, 0} +\!\!= M_{1};
&
C_{1, 1} -\!\!= M_{1};
&
C_{3, 2} +\!\!= M_{1};
&
C_{3, 3} -\!\!= M_{1}; 
\\
M_{2}  =  ( A_{0, 0} \!+\! A_{2, 2}  )  ( B_{0, 1} \!+\! B_{2, 3} \!+\! B_{1, 1} +
B_{3, 3}  );
&
  
C_{0, 1} +\!\!= M_{2};
&
C_{1, 1} +\!\!= M_{2};
&
C_{2, 3} +\!\!= M_{2};
&
C_{3, 3} +\!\!= M_{2};
\\
M_{3}  =  ( A_{1, 1} \!+\! A_{3, 3}  )  ( B_{1, 0} \!+\! B_{3, 2} \!+\! B_{0, 0} +
B_{2, 2}  );
&
  
C_{0, 0} +\!\!= M_{3};
&
C_{1, 0} +\!\!= M_{3};
&
C_{2, 2} +\!\!= M_{3};
&
C_{3, 2} +\!\!= M_{3};
\\
M_{4}  =  ( A_{0, 0} \!+\! A_{2, 2} \!+\! A_{0, 1} \!+\! A_{2, 3}  )  ( B_{1, 1} +
B_{3, 3}  );
&
  
C_{0, 0} -\!\!= M_{4};
&
C_{0, 1} +\!\!= M_{4};
&
C_{2, 2} -\!\!= M_{4};
&
C_{2, 3} +\!\!= M_{4};
\\
M_{5}  =  ( A_{1, 0} \!+\! A_{3, 2} \!+\! A_{0, 0} \!+\! A_{2, 2}  )  ( B_{0, 0} +
B_{2, 2} \!+\! B_{0, 1} \!+\! B_{2, 3}  );
&
  
C_{1, 1} +\!\!= M_{5};
&
C_{3, 3} +\!\!= M_{5};
\\
M_{6}  =  ( A_{0, 1} \!+\! A_{2, 3} \!+\! A_{1, 1} \!+\! A_{3, 3}  )  ( B_{1, 0} +
B_{3, 2} \!+\! B_{1, 1} \!+\! B_{3, 3}  );
&
  
C_{0, 0} +\!\!= M_{6};
&
C_{2, 2} +\!\!= M_{6};
\\
M_{7}  =  ( A_{2, 0} \!+\! A_{2, 2} \!+\! A_{3, 1} \!+\! A_{3, 3}  )  ( B_{0, 0} +
B_{1, 1}  );
&
  
C_{2, 0} +\!\!= M_{7};
&
C_{3, 1} +\!\!= M_{7};
&
C_{2, 2} -\!\!= M_{7};
&
C_{3, 3} -\!\!= M_{7};
\\
M_{8}  =  ( A_{3, 0} \!+\! A_{3, 2} \!+\! A_{3, 1} \!+\! A_{3, 3}  )  ( B_{0, 0}
);
&
  
C_{3, 0} +\!\!= M_{8};
&
C_{3, 1} -\!\!= M_{8};
&
C_{3, 2} -\!\!= M_{8};
&
C_{3, 3} +\!\!= M_{8};
\\
M_{9}  =  ( A_{2, 0} \!+\! A_{2, 2}  )  ( B_{0, 1} \!+\! B_{1, 1}  );
&
  
C_{2, 1} +\!\!= M_{9};
&
C_{3, 1} +\!\!= M_{9};
&
C_{2, 3} -\!\!= M_{9};
&
C_{3, 3} -\!\!= M_{9};
\\
M_{10}  =  ( A_{3, 1} \!+\! A_{3, 3}  )  ( B_{1, 0} \!+\! B_{0, 0}  );
&
  
C_{2, 0} +\!\!= M_{10};
&
C_{3, 0} +\!\!= M_{10};
&
C_{2, 2} -\!\!= M_{10};
&
C_{3, 2} -\!\!= M_{10};
\\
M_{11} = ( A_{2, 0} \!+\! A_{2, 2} \!+\! A_{2, 1} \!+\! A_{2, 3}  )  ( B_{1, 1}
); &
  
C_{2, 0} -\!\!= M_{11};
&
C_{2, 1} +\!\!= M_{11};
&
C_{2, 2} +\!\!= M_{11};
&
C_{2, 3} -\!\!= M_{11};
\\
M_{12}  =  ( A_{3, 0} \!+\! A_{3, 2} \!+\! A_{2, 0} \!+\! A_{2, 2}  )  ( B_{0, 0} +
B_{0, 1}  );
&
  
C_{3, 1} +\!\!= M_{12};
&
C_{3, 3} -\!\!= M_{12};
\\
M_{13}  =  ( A_{2, 1} \!+\! A_{2, 3} \!+\! A_{3, 1} \!+\! A_{3, 3}  )  ( B_{1, 0} +
B_{1, 1}  );
&
  
C_{2, 0} +\!\!= M_{13};
&
C_{2, 2} -\!\!= M_{13};
\\
M_{14}  =  ( A_{0, 0} \!+\! A_{1, 1}  )  ( B_{0, 2} \!+\! B_{2, 2} \!+\! B_{1, 3} +
B_{3, 3}  );
&
  
C_{0, 2} +\!\!= M_{14};
&
C_{1, 3} +\!\!= M_{14};
&
C_{2, 2} +\!\!= M_{14};
&
C_{3, 3} +\!\!= M_{14};
\\
M_{15}  =  ( A_{1, 0} \!+\! A_{1, 1}  )  ( B_{0, 2} \!+\! B_{2, 2}  );
&
  
C_{1, 2} +\!\!= M_{15};
&
C_{1, 3} -\!\!= M_{15};
&
C_{3, 2} +\!\!= M_{15};
&
C_{3, 3} -\!\!= M_{15};
\\
M_{16}  =  ( A_{0, 0}  )  ( B_{0, 3} \!+\! B_{2, 3} \!+\! B_{1, 3} \!+\! B_{3, 3}
);
&
  
C_{0, 3} +\!\!= M_{16};
&
C_{1, 3} +\!\!= M_{16};
&
C_{2, 3} +\!\!= M_{16};
&
C_{3, 3} +\!\!= M_{16};
\\
M_{17}  =  ( A_{1, 1}  )  ( B_{1, 2} \!+\! B_{3, 2} \!+\! B_{0, 2} \!+\! B_{2, 2}
);
&
  
C_{0, 2} +\!\!= M_{17};
&
C_{1, 2} +\!\!= M_{17};
&
C_{2, 2} +\!\!= M_{17};
&
C_{3, 2} +\!\!= M_{17};
\\
M_{18} = ( A_{0, 0} \!+\! A_{0, 1}  )  ( B_{1, 3} \!+\! B_{3, 3}  ); &
  
C_{0, 2} -\!\!= M_{18};
&
C_{0, 3} +\!\!= M_{18};
&
C_{2, 2} -\!\!= M_{18};
&
C_{2, 3} +\!\!= M_{18};
\\
M_{19}  =  ( A_{1, 0} \!+\! A_{0, 0}  )  ( B_{0, 2} \!+\! B_{2, 2} \!+\! B_{0, 3} +
B_{2, 3}  );
&
  
C_{1, 3} +\!\!= M_{19};
&
C_{3, 3} +\!\!= M_{19};
\\
M_{20}  =  ( A_{0, 1} \!+\! A_{1, 1}  )  ( B_{1, 2} \!+\! B_{3, 2} \!+\! B_{1, 3} +
B_{3, 3}  );
&
  
C_{0, 2} +\!\!= M_{20};
&
C_{2, 2} +\!\!= M_{20};
\\
M_{21}  =  ( A_{2, 2} \!+\! A_{3, 3}  )  ( B_{2, 0} \!+\! B_{0, 0} \!+\! B_{3, 1} +
B_{1, 1}  );
&
C_{0, 0} +\!\!= M_{21};
&
C_{1, 1} +\!\!= M_{21};
&
C_{2, 0} +\!\!= M_{21};
&
C_{3, 1} +\!\!= M_{21};
\\
M_{22}  =  ( A_{3, 2} \!+\! A_{3, 3}  )  ( B_{2, 0} \!+\! B_{0, 0}  );
&
C_{1, 0} +\!\!= M_{22};
&
C_{1, 1} -\!\!= M_{22};
&
C_{3, 0} +\!\!= M_{22};
&
C_{3, 1} -\!\!= M_{22};
\\
M_{23}  =  ( A_{2, 2}  )  ( B_{2, 1} \!+\! B_{0, 1} \!+\! B_{3, 1} \!+\! B_{1, 1}
);
&
C_{0, 1} +\!\!= M_{23};
&
C_{1, 1} +\!\!= M_{23};
&
C_{2, 1} +\!\!= M_{23};
&
C_{3, 1} +\!\!= M_{23};
\\
M_{24}  =  ( A_{3, 3}  )  ( B_{3, 0} \!+\! B_{1, 0} \!+\! B_{2, 0} \!+\! B_{0, 0}
);
&
C_{0, 0} +\!\!= M_{24};
&
C_{1, 0} +\!\!= M_{24};
&
C_{2, 0} +\!\!= M_{24};
&
C_{3, 0} +\!\!= M_{24};
\\
M_{25} = ( A_{2, 2} \!+\! A_{2, 3}  )  ( B_{3, 1} \!+\! B_{1, 1}  ); &
C_{0, 0} -\!\!= M_{25};
&
C_{0, 1} +\!\!= M_{25};
&
C_{2, 0} -\!\!= M_{25};
&
C_{2, 1} +\!\!= M_{25};
\\
M_{26}  =  ( A_{3, 2} \!+\! A_{2, 2}  )  ( B_{2, 0} \!+\! B_{0, 0} \!+\! B_{2, 1} +
B_{0, 1}  );
&
C_{1, 1} +\!\!= M_{26};
&
C_{3, 1} +\!\!= M_{26};
\\
M_{27}  =  ( A_{2, 3} \!+\! A_{3, 3}  )  ( B_{3, 0} \!+\! B_{1, 0} \!+\! B_{3, 1} +
B_{1, 1}  );
&
C_{0, 0} +\!\!= M_{27};
&
C_{2, 0} +\!\!= M_{27};
\\
M_{28} = ( A_{0, 0} \!+\! A_{0, 2} \!+\! A_{1, 1} \!+\! A_{1, 3}  )  ( B_{2, 2} +
B_{3, 3}  ); &
C_{0, 0} -\!\!= M_{28};
&
C_{1, 1} -\!\!= M_{28};
&
C_{0, 2} +\!\!= M_{28};
&
C_{1, 3} +\!\!= M_{28};
\\
M_{29} = ( A_{1, 0} \!+\! A_{1, 2} \!+\! A_{1, 1} \!+\! A_{1, 3}  )  ( B_{2, 2}
);&
C_{1, 0} -\!\!= M_{29};
&
C_{1, 1} +\!\!= M_{29};
&
C_{1, 2} +\!\!= M_{29};
&
C_{1, 3} -\!\!= M_{29};
\\
M_{30} = ( A_{0, 0} \!+\! A_{0, 2}  )  ( B_{2, 3} \!+\! B_{3, 3}  ); &
C_{0, 1} -\!\!= M_{30};
&
C_{1, 1} -\!\!= M_{30};
&
C_{0, 3} +\!\!= M_{30};
&
C_{1, 3} +\!\!= M_{30};
\\
M_{31} = ( A_{1, 1} \!+\! A_{1, 3}  )  ( B_{3, 2} \!+\! B_{2, 2}  ); &
C_{0, 0} -\!\!= M_{31};
&
C_{1, 0} -\!\!= M_{31};
&
C_{0, 2} +\!\!= M_{31};
&
C_{1, 2} +\!\!= M_{31};
\\
M_{32}  =  ( A_{0, 0} \!+\! A_{0, 2} \!+\! A_{0, 1} \!+\! A_{0, 3}  )  ( B_{3, 3}
);
&
C_{0, 0} +\!\!= M_{32};
&
C_{0, 1} -\!\!= M_{32};
&
C_{0, 2} -\!\!= M_{32};
&
C_{0, 3} +\!\!= M_{32};
\\
M_{33} = ( A_{1, 0} \!+\! A_{1, 2} \!+\! A_{0, 0} \!+\! A_{0, 2}  )  ( B_{2, 2} +
B_{2, 3}  );&
C_{1, 1} -\!\!= M_{33};
&
C_{1, 3} +\!\!= M_{33};
\\
M_{34} = ( A_{0, 1} \!+\! A_{0, 3} \!+\! A_{1, 1} \!+\! A_{1, 3}  )  ( B_{3, 2} +
B_{3, 3}  ); &
C_{0, 0} -\!\!= M_{34};
&
C_{0, 2} +\!\!= M_{34};
\\
M_{35}  =  ( A_{2, 0} \!+\! A_{0, 0} \!+\! A_{3, 1} \!+\! A_{1, 1}  )  ( B_{0, 0} +
B_{0, 2} \!+\! B_{1, 1} \!+\! B_{1, 3}  );
&
C_{2, 2} +\!\!= M_{35};
&
C_{3, 3} +\!\!= M_{35};
\\
M_{36}  =  ( A_{3, 0} \!+\! A_{1, 0} \!+\! A_{3, 1} \!+\! A_{1, 1}  )  ( B_{0, 0} +
B_{0, 2}  );
&
C_{3, 2} +\!\!= M_{36};
&
C_{3, 3} -\!\!= M_{36};
\\
M_{37}  =  ( A_{2, 0} \!+\! A_{0, 0}  )  ( B_{0, 1} \!+\! B_{0, 3} \!+\! B_{1, 1} +
B_{1, 3}  );
&
C_{2, 3} +\!\!= M_{37};
&
C_{3, 3} +\!\!= M_{37};
\\
M_{38}  =  ( A_{3, 1} \!+\! A_{1, 1}  )  ( B_{1, 0} \!+\! B_{1, 2} \!+\! B_{0, 0} +
B_{0, 2}  );
&
C_{2, 2} +\!\!= M_{38};
&
C_{3, 2} +\!\!= M_{38};
\\
M_{39} = ( A_{2, 0} \!+\! A_{0, 0} \!+\! A_{2, 1} \!+\! A_{0, 1}  )  ( B_{1, 1} +
B_{1, 3}  ); &
C_{2, 2} -\!\!= M_{39};
&
C_{2, 3} +\!\!= M_{39};
\\
M_{40}  =  ( A_{3, 0} \!+\! A_{1, 0} \!+\! A_{2, 0} \!+\! A_{0, 0}  )  ( B_{0, 0} +
B_{0, 2} \!+\! B_{0, 1} \!+\! B_{0, 3}  );
&
C_{3, 3} +\!\!= M_{40};
\\
M_{41}  =  ( A_{2, 1} \!+\! A_{0, 1} \!+\! A_{3, 1} \!+\! A_{1, 1}  )  ( B_{1, 0} +
B_{1, 2} \!+\! B_{1, 1} \!+\! B_{1, 3}  );
&
C_{2, 2} +\!\!= M_{41};
\\
M_{42}  =  ( A_{0, 2} \!+\! A_{2, 2} \!+\! A_{1, 3} \!+\! A_{3, 3}  )  ( B_{2, 0} +
B_{2, 2} \!+\! B_{3, 1} \!+\! B_{3, 3}  );
&
C_{0, 0} +\!\!= M_{42};
&
C_{1, 1} +\!\!= M_{42};
\\
M_{43}  =  ( A_{1, 2} \!+\! A_{3, 2} \!+\! A_{1, 3} \!+\! A_{3, 3}  )  ( B_{2, 0} +
B_{2, 2}  );
&
C_{1, 0} +\!\!= M_{43};
&
C_{1, 1} -\!\!= M_{43};
\\
M_{44}  =  ( A_{0, 2} \!+\! A_{2, 2}  )  ( B_{2, 1} \!+\! B_{2, 3} \!+\! B_{3, 1} +
B_{3, 3}  );
&
C_{0, 1} +\!\!= M_{44};
&
C_{1, 1} +\!\!= M_{44};
\\
M_{45}  =  ( A_{1, 3} \!+\! A_{3, 3}  )  ( B_{3, 0} \!+\! B_{3, 2} \!+\! B_{2, 0} +
B_{2, 2}  );
&
C_{0, 0} +\!\!= M_{45};
&
C_{1, 0} +\!\!= M_{45};
\\
M_{46} = ( A_{0, 2} \!+\! A_{2, 2} \!+\! A_{0, 3} \!+\! A_{2, 3}  )  ( B_{3, 1} +
B_{3, 3}  ); &
C_{0, 0} -\!\!= M_{46};
&
C_{0, 1} +\!\!= M_{46};
\\
M_{47}  =  ( A_{1, 2} \!+\! A_{3, 2} \!+\! A_{0, 2} \!+\! A_{2, 2}  )  ( B_{2, 0} +
B_{2, 2} \!+\! B_{2, 1} \!+\! B_{2, 3}  );
&
C_{1, 1} +\!\!= M_{47};
\\
M_{48}  =  ( A_{0, 3} \!+\! A_{2, 3} \!+\! A_{1, 3} \!+\! A_{3, 3}  )  ( B_{3, 0} +
B_{3, 2} \!+\! B_{3, 1} \!+\! B_{3, 3}  );
&
  C_{0, 0} +\!\!= M_{48}; \\
\end{array}
\\
\hline
\end{array}
\]%
}
\caption{Computations for two-level \strassen{}.}
\label{fig:2stepB}
\end{figure*}

%% file: 04analysis.tex
\fromto{We have shown that all operations of any level of \strassen\ can be cast into a combination of general operations (\ref{e:straprim4}).}{}
\fromto{In Section \ref{s:stra}, w}{W}e \fromto{}{now} discuss the \fromto{optimization}{} details of how we adapt the high-performance\fromto{GEMM}{} \GOTO\ approach \fromto{(Section \ref{s:gemm}) for}{to} these \fromto{general}{specialized} operations \fromto{as the}{to yield} building blocks for \fromto{reloaded}{a family of} \strassen\ \fromto{algorithm (Section \ref{s:reload})}{implementations}. \fromto{Further, in Section \ref{s:var}, we develop various implementations to evaluate the benefits of different technique we proposed. Finally, w}{Next, w}e \fromto{introduce}{also give} a \fromto{theoretical}{} performance model \fromto{in Section \ref{s:model}}{} for comparing \fromto{implementation variations of \strassen\ and provide an analytical method for tuning reloaded \strassen{}}{members of this family}.

\NoShow{\subsection{Adapting the \GOTO{} approach for \strassen{}} \label{s:stra}
\input stra.tex

}

\subsection{Implementations} \label{s:var}
\input implementation

\subsection{Performance Model} \label{s:model}

\input model

%% file: stra.tex
We present a \fromto{new}{family of} implementation\fromto{}{s} of \strassen, by incorporating the additional memory movement into the inherent packing \fromto{buffer}{that happens} inside \fromto{a high-performance GEMM}{the BLIS framework}, and\fromto{}{/or}  
fusing the GEMM \fromto{}{micro}kernel with the updating of multiple submatrices of the \fromto{product}{result} matrix $C$.
\fromto{We expose the benefits of utilizing the packing buffer by keeping it in the fast cache (\figref{fig:side_by_side}(right)), 
and reusing the intermediate output $M$ (
\figref{fig:strassenlevel1}) immediately after it is computed in the kernel, so less memory latency is suffered. To increase the L1 cache resuse rate, updating of the
submatrices of $C$ is performed as soon as a small block of $M$ has been computed, because $M$ resides in register at that time.
To achieve that we need to adapt the \GOTO{} algorithm under the hood of the modern high-performance implementation of GEMM.
}{}
Here we use the general operation (\ref{e:straprim1}) for one-level Strassen as an example to \fromto{show}{discuss} how we implement \fromto{a high-performance practical \strassen{} with \GOTO{} approach.}{members of the family of implementations.}
As discussed in the previous section, the implementations for other general operations of any level \strassen\ are similar.
\fromto{}{In our discussion we refer to the loop structure given in \figref{fig:side_by_side}.}

\fromto{We first present the pseudo-code of our implementation of the general operation (\ref{e:straprim1}) in \figref{fig:blis_strassen1}. We later discuss the details about how we incorporate the
packing routines, how we design the micro-kernel, how we select the parameters for various architectures, and how we parallelize our \strassen{} in a simple but effective way.}{}

\NoShow{\begin{figure}[tb!]
    \begin{center}
         \begin{minipage}[t]{5in}
             \footnotesize
             \mbox{\input blis_strassen1 }
         \end{minipage}
     \end{center}
     \caption{Implementation of general operation (\ref{e:straprim1}), mapped to \figref{fig:side_by_side}(right).}
 \label{fig:blis_strassen1}
\end{figure}

\subsubsection*{Loop structure}
Our implementation is composed of six layers of loops, related to different partitioning along $m$, $k$ and $n$ dimensions. The partition scheme efficiently amortizes memory blocking and alignment to achieve high reuse efficiency.
We suggest the reader check both \figref{fig:side_by_side}(right) and \figref{fig:blis_strassen1} to better understand the details. In the outer-most loop (Loop 5: indexed by $j_C$), both $V$ and $W$ (submatrices of $A$) are partitioned with block size $n_C$ along $n$ dimension. Loop 4 (indexed by $p_C$) partitions the $k$ dimension of submatrices of both $A$ and $B$ (i.e. $X$, $Y$, $V$ and $W$) with the block size $k_C$. Loop 3 (indexed by $i_C$) further partitions the $m$ dimension of both $X$ and $Y$ (submatrices of $B$) with block size $m_C$.
Note that we form new buffers $ \widetilde B_p $ and $ \widetilde A_i $ before and after Loop 3, separately. We will explain this later.
All loops inside Loop 2 (including Loop 2) comprises the macro-kernel. Loop 2 and Loop 1 (indexed by $j_R$ and $i_R$) further partitions $ \widetilde B_p $ and $ \widetilde A_i $ into small tiles.
Finally, Loop 0 (indexed by $p_r$) forms the micro-kernel, which computes the intermediate result $M$ with size $m_R\times n_R$ in the registers. $C$ and $D$ will be immediately updated, as soon as Loop 0 is finished.
}

\begin{center}
	\color{red}
	I have deleted a lot...
\end{center}

\NoShow{
	\subsubsection*{Packing}

{We will see next is that by incorporating the addition of matrices in the packing and updating $ C $ and $ D $ simultaneously, essentially no temporary storage beyond the buffers already incorporated in the standard \BLIS{} approach for \gemm\     is needed.  This eliminates the explicit addition of submatrices and the data movement between memory layers that is associated with such additions.}

\subsubsection*{Micro-kernel}

Micro-kernel storing

\subsubsection*{Parameter selection}

We need to select the architecture dependent parameters $n_C$, $k_C$, $m_C$, $n_R$, $m_R$.

Adopting the method in \cite{BLIS-Analytical}, we demonstrate how to choose the parameters on Intel Xeon Phi architecture.

$n_R$ and $m_R$

\subsubsection*{Parallelization}
Our implementation incorporates the same parallelism

orthogonal to task parallelism

Orthogonal way




\subsubsection*{Portibility}
The memory packing scheme in the traditional GEMM is portable to multi-core and many-core architecture. Plus, we are adapting \GOTO{} with \BLIS{} framework, which reduces the architecture dependent part to only the micro-kernel level. We have \strassen{} implementations on a x86 architecture (multi-core) and Xeon Phi (many-core). We can further port our approach to 
BG/Q, Power7, Cortex/A8, Loonsgon/3A, etc in a similar way as \BLIS{}\cite{BLIS2}.
}

%% file: blis_strassen1.tex
\begin{tabular}{l@{\hspace{6pt}}l@{}}
\cblue{Loop 5} &{\bf for} $j_c\!=\! 0:n\!-\!1$ {\bf steps of} $n_c$ \\
& \hspace{2ex}  $\Jc\!=\! j_c:j_c\!+\!n_c\!-\!1$\\
\cblue{Loop 4} & \hspace{2ex}  {\bf for} $p_c \!=\! 0:k\!-\!1$ {\bf steps of}
         $k_c$ \\
& \hspace{4ex} $\Pc \!=\! p_c:p_c\!+\!k_c\!-\!1$\\
&\hspace{4ex}           \textcolor{black}{$V(\Pc,\Jc)+ \epsilon W(\Pc,\Jc)$} $\rightarrow \textcolor{black}{ \widetilde B_p }$  \myalgcc{Pack into $ \widetilde B_p $}\\
\cblue{Loop 3} & \hspace{4ex}           {\bf for} $i_c \!=\! 0:m\!-\!1$ {\bf
         steps of} $m_c$ \\
& \hspace{6ex} $\Ic \!=\! i_c:i_c\!+\!m_c\!-\!1$\\
&\hspace{6ex}                     \textcolor{black}{$X(\Ic,\Pc)+ \delta Y(\Ic,\Pc)$} $\rightarrow \textcolor{black}{ \widetilde A_i }$  \myalgcc{Pack into $  \widetilde A_i $} \\ \cline{2-2} 
& \hspace{6ex} // {\tt macro-kernel}\\ 
\cblue{Loop 2} &\hspace{6ex} {\bf for} $j_r \!=\! 0:n_c\!-\!1$ {\bf steps of}
         $n_r$ \\
& \hspace{8ex}  $\Jr \!=\! j_r:j_r\!\!+\!\!n_r\!-\!1$  \\
\cblue{Loop 1} &\hspace{8ex}   {\bf for} $i_r \!=\! 0:m_c\!-\!1$ {\bf  steps
        of} $m_r$ \\
& \hspace{10ex}  $\Ir \!=\! i_r:i_r\!+\!m_r\!-\!1$\\
\cline{2-2} 
&\hspace{10ex} \cgreen{//{\tt micro-kernel}} \\
\cblue{Loop 0} &\hspace{10ex} {\bf for} $p_r\!=\! 0:p_c\!-\!1$ {\bf steps of} $1$ \\
&\hspace{12ex} \textcolor{black}{$M_r(\Ir,\Jr)$}~$\mathrel{\!+\!}=$~\textcolor{black}{${\widetilde A_i}(\Ir,p_r)$}~\textcolor{black}{$ {\widetilde B_p}(p_r,\Jr)$} \\
&\hspace{10ex} {\bf endfor}\\
&\hspace{10ex} \textcolor{black}{$C(\Ir+i_c,\Jr+j_c)$}~$\mathrel{\!+\!}=$~\textcolor{black}{$\alpha \gamma_0 M_r(\Ir,\Jr)$}\\
&\hspace{10ex} \textcolor{black}{$D(\Ir+i_c,\Jr+j_c)$}~$\mathrel{\!+\!}=$~\textcolor{black}{$\alpha \gamma_1 M_r(\Ir,\Jr)$}\\
\cline{2-2} 
&\hspace{8ex} {\bf endfor}\\
&\hspace{4ex} {\bf endfor}\\
\cline{2-2} 
&\hspace{2ex} {\bf endfor}\\ 
&{\bf endfor}\\ 
\end{tabular}

%% file: implementation.tex
We implement \fromto{the following variations}{a family of algorithms} for \fromto{both one-level and two-level \strassen{} algorithms.}{up to two levels of \strassen,}
\fromto{These allow us to assess the benefits of the different techniques we implement in Section \ref{s:stra}.
For our implementations, components from the \BLIS\ framework were modified.  We refer to the submatrices in \figref{fig:side_by_side} as we describe these.}{building upon the \BLIS\ framework.}

\subsubsection*{Building blocks}

The \BLIS\ framework provides three primitives for composing \dgemm:
a routine for packing $ B_p $ into $ \widetilde{B}_p$, 
a routine for packing $ A_i $ into $ \widetilde A_i $, 
and a micro-kernel for updating an $ m_R \times n_R $ submatrix of $ C $.
The first two are typically written in {\tt C} while the last one is typically written in (inlined) assembly code.

To implement a typical operation given in (\ref{e:straprim4}), 
\begin{itemize}
	\item 
the routine for packing $ B_p $ is modified to integrate the addition of  multiple matrices $ V_t $ into packed buffer $ \widetilde B_p $;
\item
the routine for packing $ A_i $ is modified to integrate the addition of  multiple matrices $ X_s $ into packed buffer $ \widetilde A_i $; and
\item
the micro-kernel is modified to integrate the addition of the result to multiple submatrices.
\end{itemize}

\subsubsection*{Variations on a theme}
The members of our family of \strassen\ implementations differ by 
how many levels of \strassen\ they incorporate and 
which of the above described modified primitives they use:
\begin{itemize}
	\item 
	\XXXstrassen: A traditional implementation with temporary buffers.
	\item
	\ABXstrassen: Integrates the addition of matrices into the packing of buffers $ \widetilde A_i $ and $ \widetilde B_p $ but creates explicit temporary buffers for matrices $ M $.
	\item
	\ABCstrassen: Integrates the addition of matrices into the packing of buffers $ \widetilde A_i $ and $ \widetilde B_p $ {\em and} the addition of the result of the micro-kernel computation to multiple submatrices of $ C $.
	For small problem size $ k $ this version has the advantage over \ABXstrassen\ that the temporary matrix $ M $ is not moved in and out of memory multiple times.  The disadvantage is that for large $k $ the submatrices of $ C $ to which contributions are added are moved in and out of memory multiple times instead.
\end{itemize}

%% file: model.tex
In order to compare the performance of the traditional BLAS \dgemm\ routine and the various implementations of
\strassen, we define the \emph{effective} \GFLOPS{} metric for
$m\times k \times n$ matrix multiplication, similar to \cite{StrassenBenson,PPL2,StrassenLipshitz}:
\begin{equation}
  \text{\emph{effective} GFLOPS} = \frac{2\cdot m\cdot n\cdot k}{\text{time (in seconds)}}\cdot 10 ^{-9}.
  \label{e:egflops1}
\end{equation}
We next derive a model to predict the execution time $T$ and the \emph{effective} \GFLOPS\ of the traditional
BLAS \dgemm\ and the various implementations of \strassen.  
Theoretical predictions allow us to compare and contrast different implementation decisions, help with 
performance debugging, and (if sufficiently accurate) can be used to 
choose the right member of the family of implementations as a function of the number of threads used and/or problem size.

\begin{figure}[!t]
\centering
{\footnotesize
  \begin{tabular}{l| ccrrr}
  \whline
                                        & type
                                        & $\tau$
                                        & \dgemm{}
                                        & one-level 
                                        & two-level  \\
  \hline
  $T_{a}^{\times}$                      & -
                                        & $\tau_{a}$
                                        & $ 2 m  n  k $
                                        & $7\times2\frac{m}{2}\frac{n}{2}\frac{k}{2}$
                                        & $49\times2\frac{m}{4}\frac{n}{4}\frac{k}{4}$ \\
  $T_{a}^{A_{+}}$                           & -
                                        & $\tau_{a}$
                                        & -
                                        & $5\times2\frac{m}{2} \frac{k}{2}$
                                        & $95\times2\frac{m}{4}\frac{k}{4}$ \\
  $T_{a}^{B_{+}}$                           & -
                                        & $\tau_{a}$
                                        & -
                                        & $5\times2\frac{k}{2} \frac{n}{2}$
                                        & $95\times2\frac{k}{4}\frac{n}{4}$ \\
  $T_{a}^{C_{+}}$                           & -
                                        & $\tau_{a}$
                                        & -
                                        & $12\times2\frac{m}{2}\frac{n}{2}$
                                        & $154\times2\frac{m}{4}\frac{n}{4}$ \\
  \hline
  $T_{m}^{A_{\times}}$                  & \texttt{r}
                                        & $\tau_{b}$
                                        & $mk \lceil \frac{n}{n_c} \rceil$
                                        & $\frac{m}{2} \frac{k}{2} \lceil \frac{n/2}{n_c} \rceil$
                                        & $\frac{m}{4} \frac{k}{4} \lceil \frac{n/4}{n_c} \rceil$ \\
  $T_{m}^{{\widetilde A}_{\times}}$     & \texttt{w}
                                        & $\tau_{b}$
                                        & $mk \lceil \frac{n}{n_c} \rceil$
                                        & $\frac{m}{2} \frac{k}{2} \lceil \frac{n/2}{n_c} \rceil$
                                        & $\frac{m}{4} \frac{k}{4} \lceil \frac{n/4}{n_c} \rceil$ \\
  $T_{m}^{B_{\times}}$                  & \texttt{r}
                                        & $\tau_{b}$
                                        & $nk$
                                        & $\frac{n}{2} \frac{k}{2}$
                                        & $\frac{n}{4} \frac{k}{4}$ \\
  $T_{m}^{{\widetilde B}_{\times}}$       & \texttt{w}
                                        & $\tau_{b}$
                                        & $nk$
                                        & $\frac{n}{2} \frac{k}{2}$
                                        & $\frac{n}{4} \frac{k}{4}$ \\
  $T_{m}^{C_{\times}}$ \fromto{}{(*)}                 & \texttt{r/w}
                                        & $\tau_{b}$
                                        & $2mn\lceil\frac{k}{k_c}\rceil$
                                        & $2\frac{m}{2}\frac{n}{2}\lceil\frac{k/2}{k_c}\rceil $
                                        & $2\frac{m}{4}\frac{n}{4}\lceil\frac{k/4}{k_c}\rceil $ \\
  \hline
  $T_{m}^{A_{+}}$                       & \texttt{r/w}
                                        & $\tau_{b}$
                                        & $mk$
                                        & $\frac{m}{2} \frac{k}{2}$
                                        & $\frac{m}{4} \frac{k}{4}$ \\
  $T_{m}^{B_{+}}$                       & \texttt{r/w}
                                        & $\tau_{b}$
                                        & $nk$
                                        & $\frac{n}{2} \frac{k}{2}$
                                        & $\frac{n}{4} \frac{k}{4}$ \\
  $T_{m}^{C_{+}}$                       & \texttt{r/w}
                                        & $\tau_{b}$
                                        & $mn$
                                        & $\frac{m}{2} \frac{n}{2}$
                                        & $\frac{m}{4} \frac{n}{4}$ \\

  \whline
  \end{tabular}
}

\vspace{0.2in}

\centering
{\footnotesize
  \begin{tabular}{l  c | rrrrrr}
  \whline
  & & $N_{m}^{A_{\times}}$  &    $N_{m}^{B_{\times}}$  &    $N_{m}^{C_{\times}}$  &    $N_{m}^{A_{+}}$  &    $N_{m}^{B_{+}}$  & $N_{m}^{C_{+}}$ \\
\hline
                    \dgemm\ &        & 1    & 1     & 1     & -    & -     & -     \\
\hline  
\multirow{ 3 }{*}{one-level}  & ABC    & 12   & 12    & 12    & -    & -     & -     \\
                            & AB     & 12   & 12    & 7     & -    & -     & 36    \\
                            & Naive  & 7    & 7     & 7     & 19   & 19    & 36    \\
\hline                                          
\multirow{ 3 }{*}{two-level}  & ABC    & 194  & 194   & 154   & -    & -     & -     \\
                            & AB     & 194  & 194   & 49    & -    & -     & 462   \\
                            & Naive  & 49   & 49    & 49    & 293  & 293   & 462   \\
  \whline
  \end{tabular}
}
\caption{
The top table shows theoretical run time breakdown analysis of BLAS \dgemm{} and various implementations of \strassen{}.
The time shown in the first column for \dgemm{},
one-level \strassen{}, two-level \strassen{}
can be computed separately by multiplying the parameter in $\tau$ column with the number in the corresponding entries.
\fromto{}{{Due to the software prefetching effects, the row marked with $(*)$ needs to be multiplied by an additional parameter $\lambda\in[0.5,1]$, which denotes
the prefetching efficiency.}}
The bottom table shows
the coefficient mapping table for computing $T_m$ in the performance model.
}
\label{tab:breakdown}
\end{figure}

\begin{figure*}[htp!]
	\centering
	\includegraphics[width=0.43\textwidth]{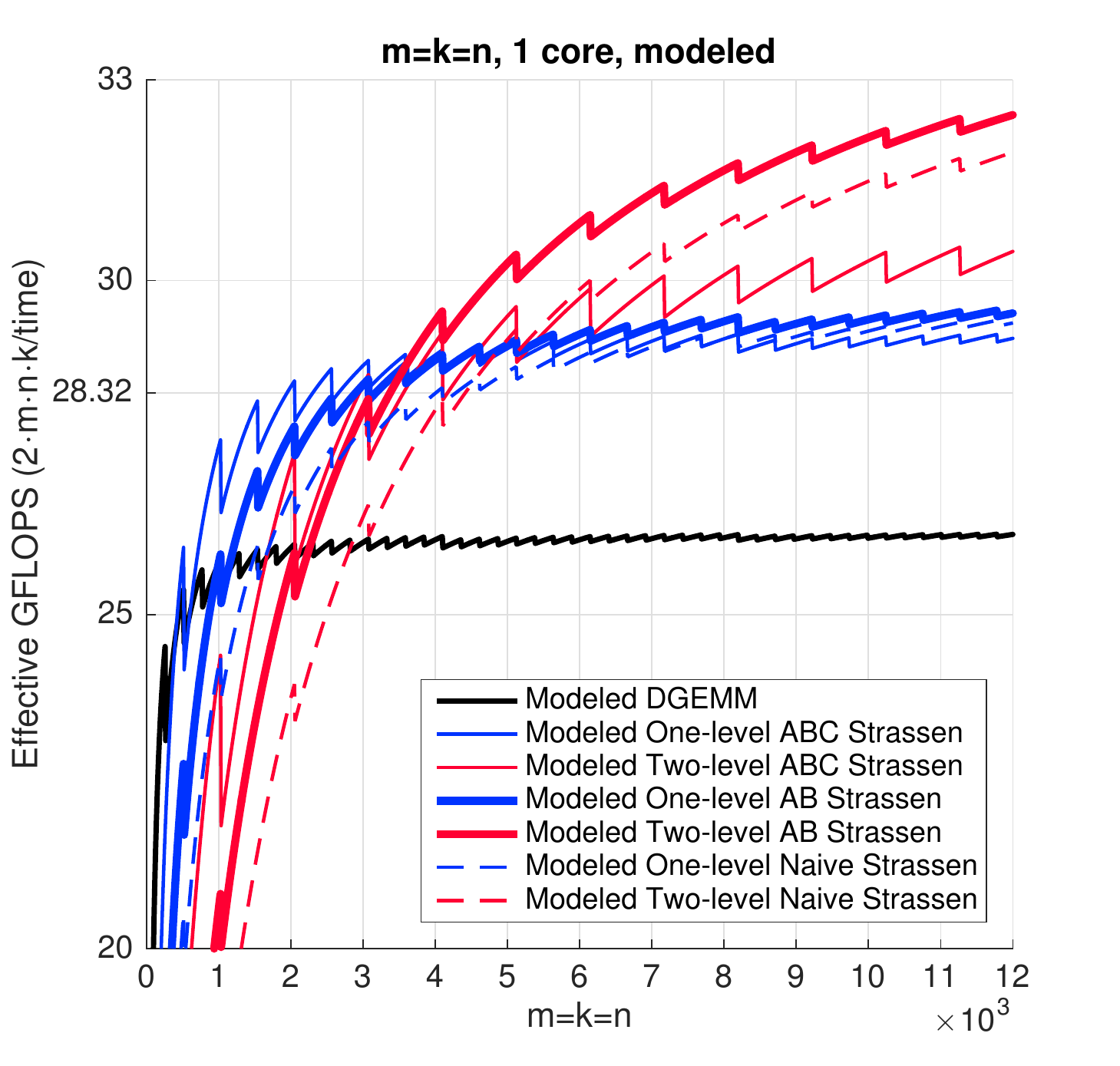}
    \includegraphics[width=0.43\textwidth]{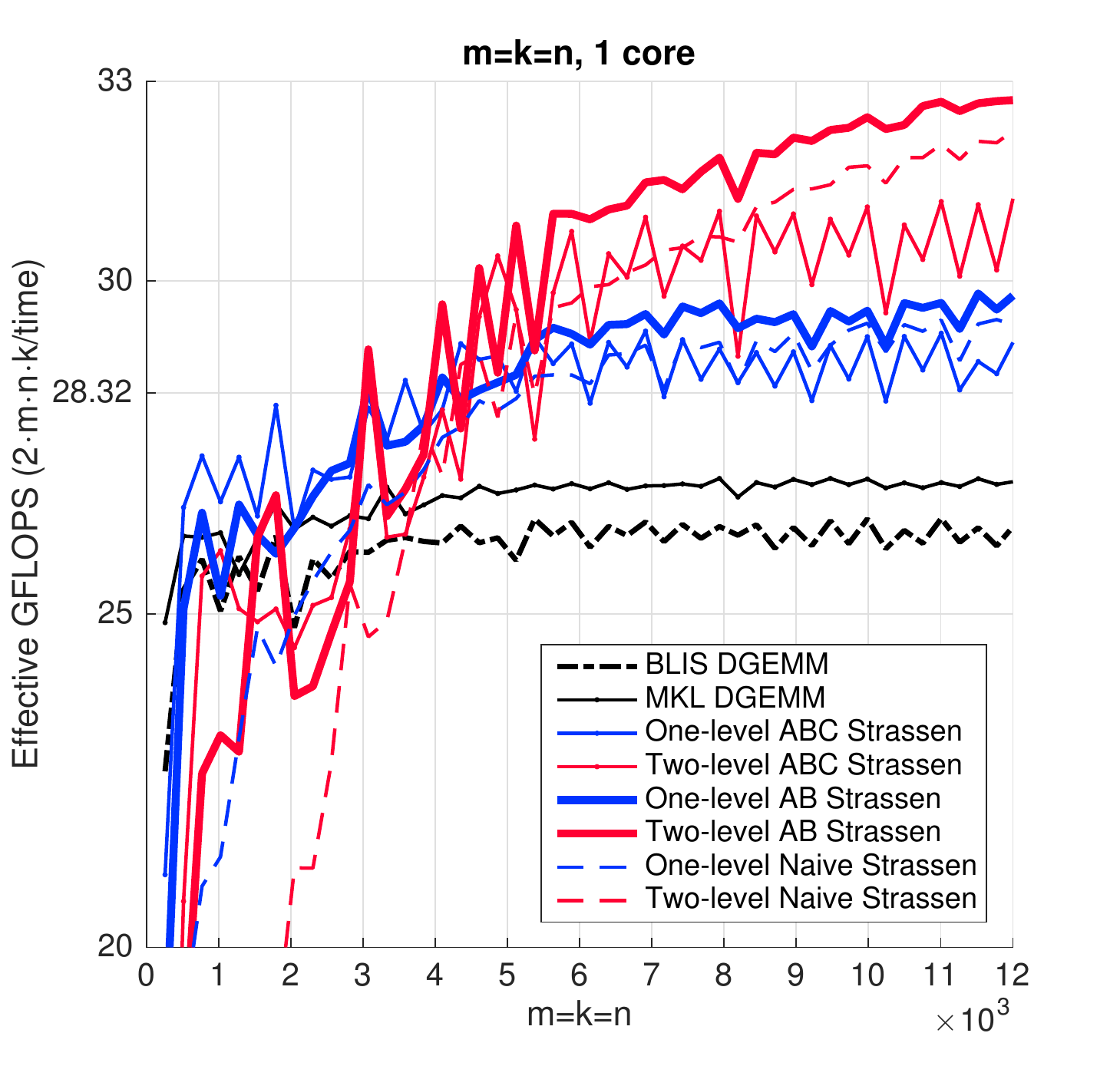}\\
    \includegraphics[width=0.43\textwidth]{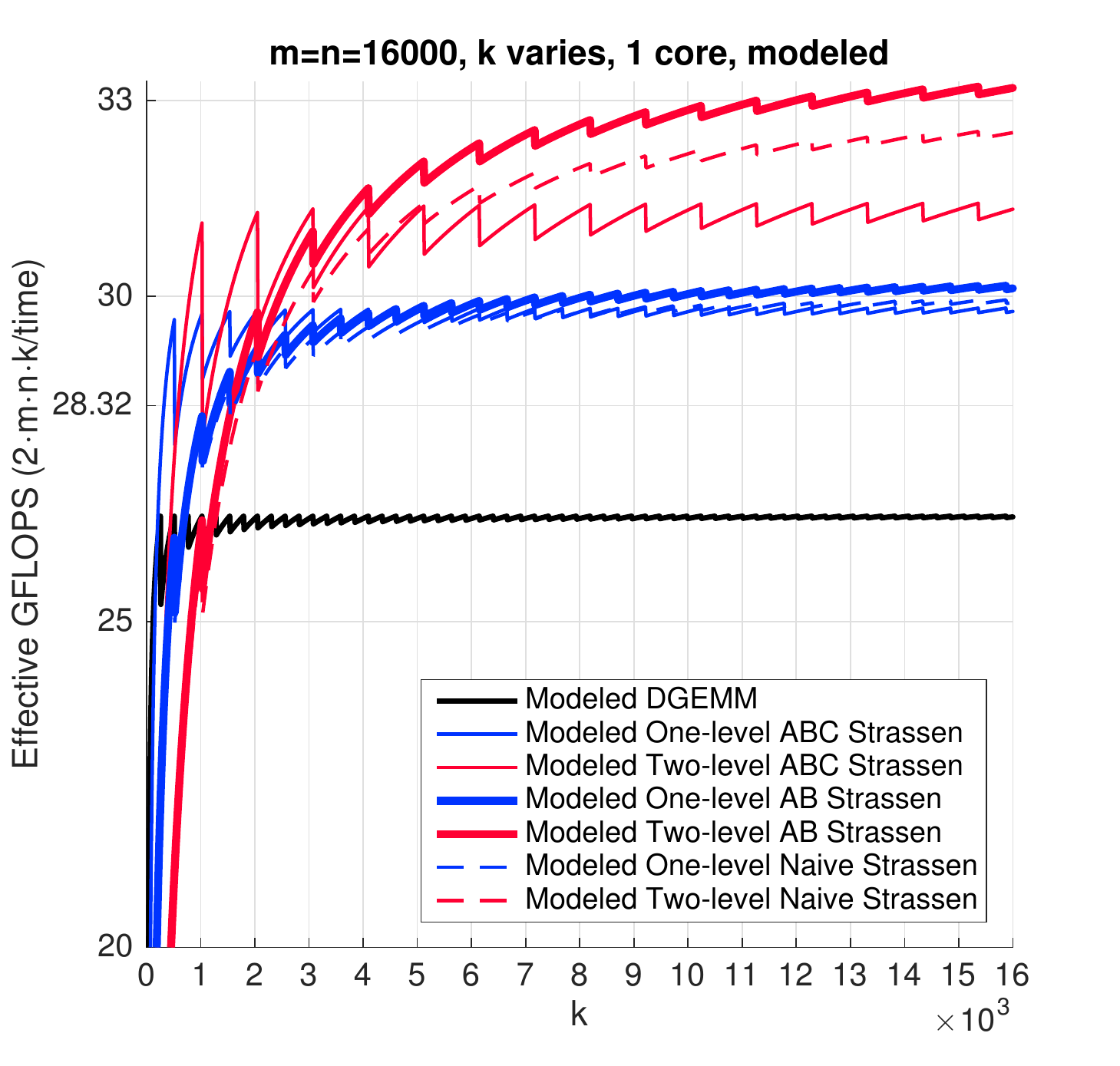}
    \includegraphics[width=0.43\textwidth]{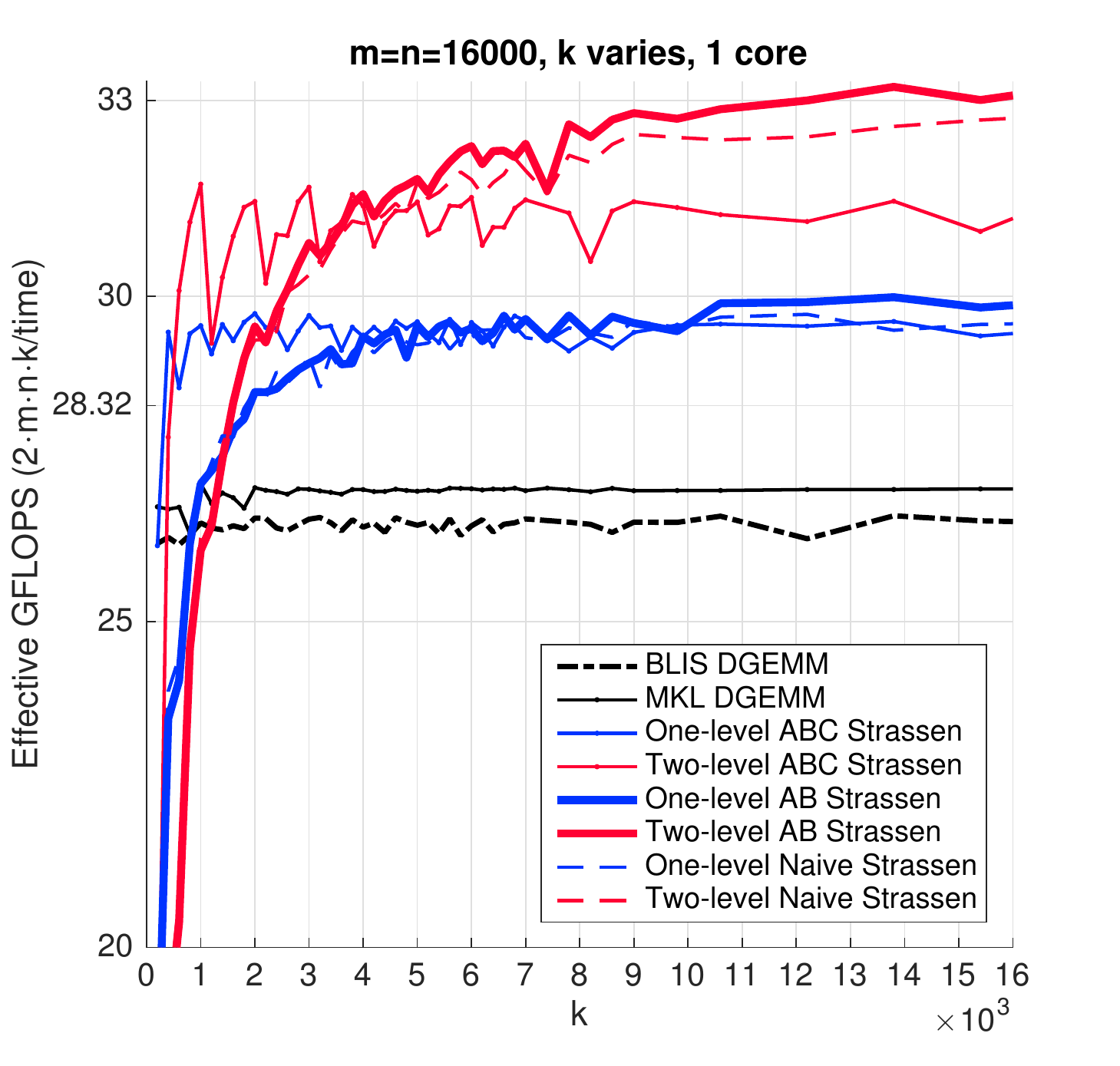}
    \includegraphics[width=0.43\textwidth]{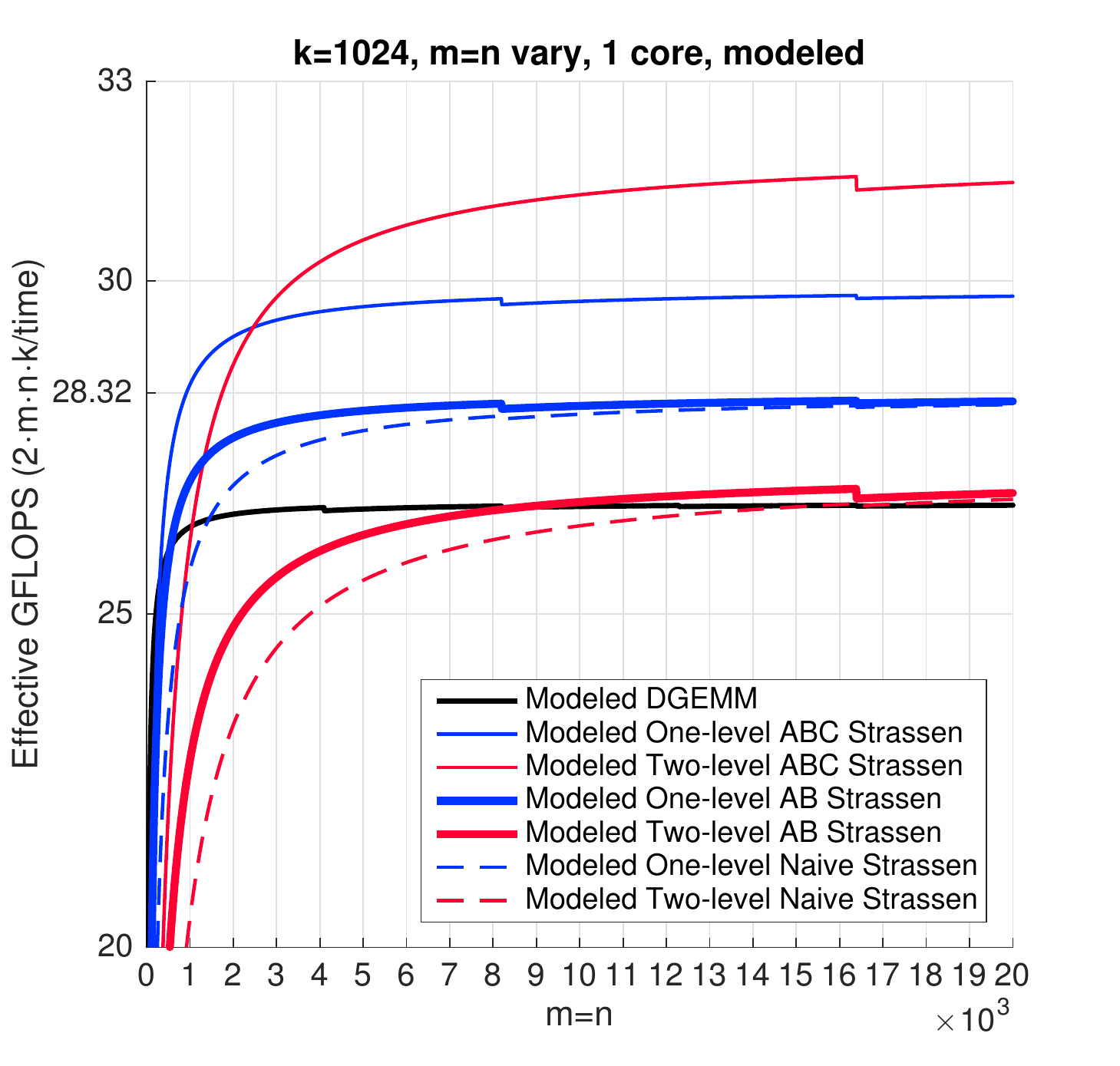}
    \includegraphics[width=0.43\textwidth]{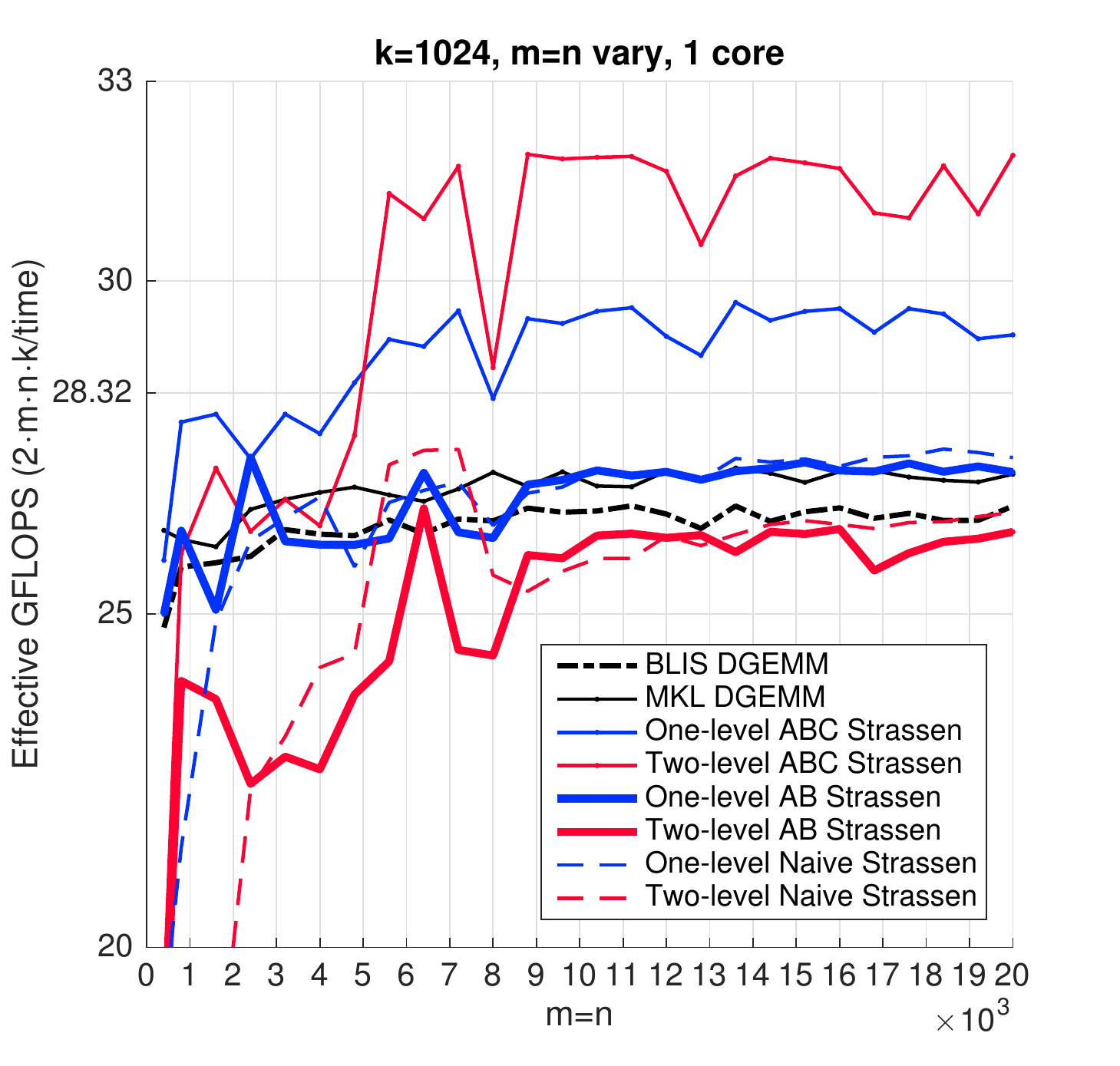}
    \NoShow{\\
	\includegraphics[width=0.43\textwidth]{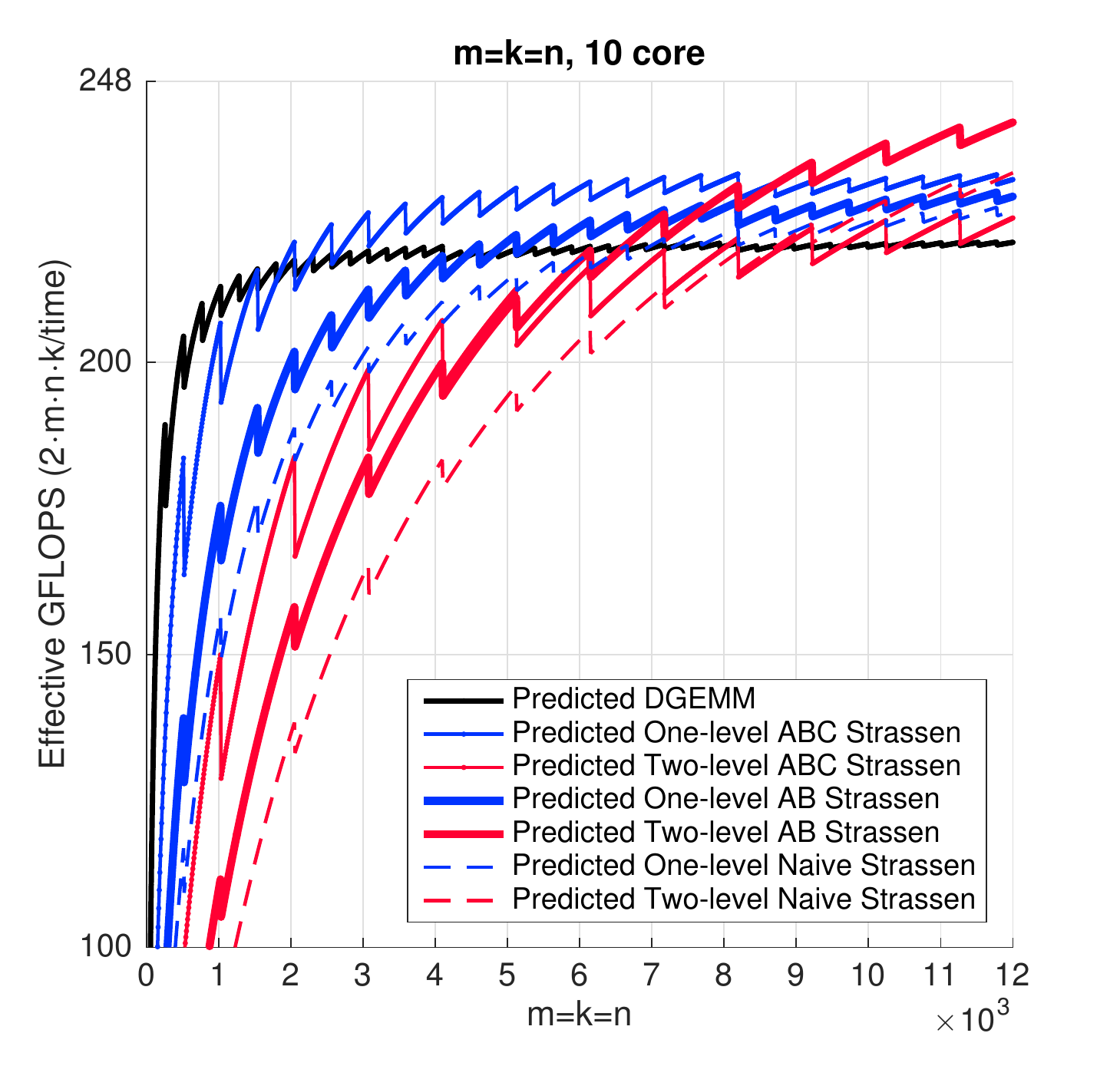}
		\includegraphics[width=0.43\textwidth]{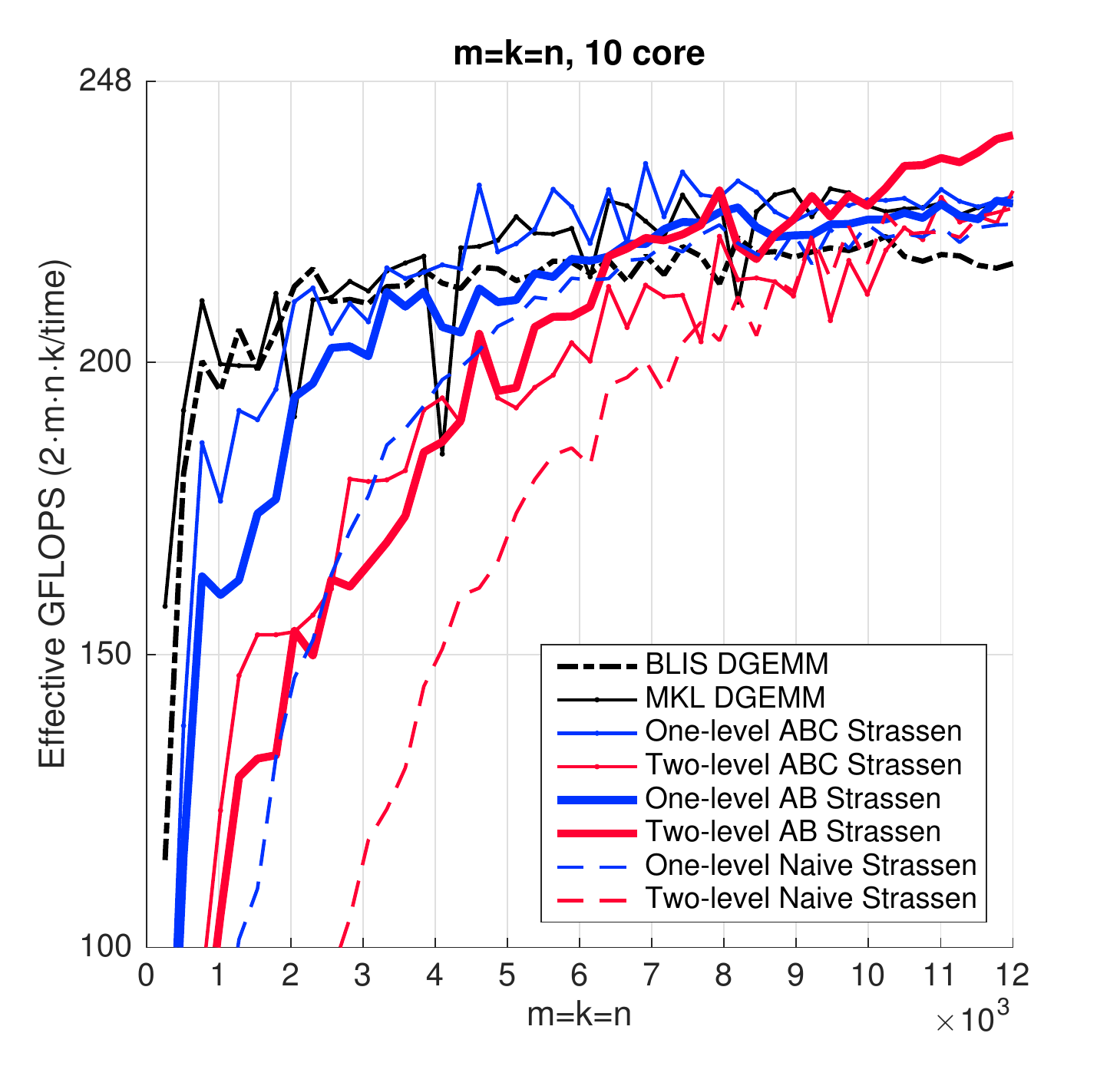}
	}
	\caption{Performance of the various implementations on an Intel\textregistered\ Xeon\textregistered\ E5 2680 v2 (Ivybridge) processor.
		Left: modeled performance. Right: actual performance.
		The range of the y-axis does not start at zero to make the graphs more readable and 28.32 marks theoretical peak performance for this architecture.
	}
	\label{fig:model}
\end{figure*} 

\subsubsection*{Assumption}
Our performance model assumes that the underlying architecture has a modern \fromto{cache}{memory} hierarchy \fromto{with 
a small piece of fast memory (L1/L2/LLC cache) and a large chunk of slow memory (DRAM).}{with fast caches and relatively slow main memory (DRAM).}
We\fromto{ also}{} assume the latency for accessing the fast memory can be ignored \fromto{}{(either because it can be overlapped with computation or because it can be amortized over sufficient computation)} while the latency of loading from main memory is exposed.
For memory store operations, our model assumes that a lazy write-back policy guarantees the time for storing
into fast memory can be hidden.
The slow memory operations for BLAS \dgemm\ and the various implementation of \strassen{} consist of three
parts: (1) memory packing in (adapted) \dgemm\ routine; (2) reading/writing the submatrices of $C$ in (adapted) \dgemm\ routine; and (3) reading/writing of
the temporary buffer that are part of  \XXXstrassen\ and \ABXstrassen, outside (adapted) \dgemm{} routine.
Based on these assumptions, the execution time is dominated by the arithmetic operations and the slow memory operations.

\subsubsection*{Notation}
\fromto{We define $\tau_{a}$, $\tau_{b}$, $T_{a}$, $T_{m}$ and $T$ here.}{}
Parameter $\tau_{a}$  denotes the time (in seconds) of one \underline{a}rithmetic (floating point) operation, i.e., the reciprocal of the theoretical peak \GFLOPS\ of the system.
Parameter $\tau_{b}$ (\underline{b}andwidth, memory operation) denotes the amortized time (in seconds) of a unit (one double precision floating point number,
or eight bytes) of contiguous data movement from DRAM to cache.
\fromto{}{{In practice, \[\tau_{b}=\frac{8 \mbox{(Bytes)}}{\mbox{bandwidth (in GBytes/s)}}\cdot 10^{-9}.\]
For single core, we need to further multiply it by the number of channels.}}

The total execution time (in seconds), $ T $, is broken down into 
the time for \underline{a}rithmetic operations, $T_{a}$, and
\underline{m}emory operations:
\begin{equation}
  T=T_{a}+T_{m}.
  \label{e:etotal}
\end{equation}
\NoShow{
With this, \eref{e:egflops1} can be rewritten as
\begin{equation}
  \text{\emph{effective} GFLOPS} = \frac{2\cdot m\cdot n\cdot k}{T_{a}+T_{m}}\cdot 10 ^{-9}.
  \label{e:egflops2}
\end{equation}
}


\subsubsection*{Arithmetic Operations}
We break down $T_{a}$ into separate terms: 
\begin{equation}
  T_{a} = T_{a}^{\times} + T_{a}^{A_{+}} + T_{a}^{B_{+}} + T_{a}^{C_{+}}, 
  \label{e:fptime}
\end{equation}
where $T_{a}^{\times}$ is the arithmetic time for submatrix multiplication, and $T_{a}^{A_{+}}$, $T_{a}^{B_{+}}$, $T_{a}^{C_{+}}$ 
denote the arithmetic time of extra additions with submatrices of $A$, $B$, $C$, respectively. For \dgemm\, since there are no
extra additions, $T_{a} = 2mnk \cdot\tau_{a}$. For one-level \strassen{}, $T_a$ is comprised of 7 submatrix multiplications, 5 extra additions
of submatrices of $A$ and $B$, and 12 extra additions of submatrices of $C$. Therefore, 
\fromto{$T_{a} = \tau_{a}\cdot(1.75mnk+2.5mk+2.5kn+6mn) $}{$T_{a} = (1.75mnk+2.5mk+2.5kn+6mn) \cdot \tau_{a} $}.
\fromto{\color{red} I prefer the $ \tau_a $ etc. at the end of the formula.}{}
Note that the matrix addition actually involves 2 floating point operations for each entry because they are cast as\fromto{``{\tt axpy}'' or bottom level}{}
{\tt FMA} instructions.
Similar analyses can be applied to compute $T_{a}$ of a two-level \strassen\ implementation.
A full analysis is summarized in \figref{tab:breakdown}.

\subsubsection*{Memory Operations}
The total data movement overhead is determined by both the original matrix sizes $m$, $n$, $k$, and block sizes
$m_C$, $n_C$, $k_C$ in our implementation \figref{fig:side_by_side}(right).
We characterize each memory operation term in \figref{tab:breakdown} by its read/write type and the amount of memory (one unit$=$double precision floating number size$=$eight bytes) involved in the movement.
We decompose $T_{m}$ into
\begin{multline}
  T_{m} = N_{m}^{A_{\times}} \cdot T_{m}^{A_{\times}} + N_{m}^{B_{\times}} \cdot T_{m}^{B_{\times}} + N_{m}^{C_{\times}} \cdot T_{m}^{C_{\times}}  
  + N_{m}^{A_{+}} \cdot T_{m}^{A_{+}} + N_{m}^{B_{+}} \cdot T_{m}^{B_{+}} + N_{m}^{C_{+}} \cdot T_{m}^{C_{+}}, 
\end{multline}
where $T_{m}^{A_{\times}}$, $T_{m}^{B_{\times}}$ are the data movement time for reading from submatrices of $A$, $B$, respectively, for packing
 inside \GOTO{} \gemm\ algorithm (\figref{fig:side_by_side});
$T_{m}^{C_{\times}}$ is the data movement time for loading \emph{and} storing submatrices of $C$ inside \gemm\ algorithm;
$T_{m}^{A_{+}}$, $T_{m}^{B_{+}}$, $T_{m}^{C_{+}}$ are the data movement time for loading \emph{or} storing submatrices of $A$, $B$, $C$, respectively, related to the temporary buffer as part of \XXXstrassen\ and \ABXstrassen{}; the $N_m^X$s denote the corresponding coefficients, which are also tabulated in \figref{tab:breakdown}.

All write operations ($T_{m}^{{\widetilde A}_{\times}}$, $T_{m}^{{\widetilde B}_{\times}}$ for storing submatrices of $A$, $B$, respectively, into packing buffers) are omitted because our assumption of lazy write-back policy with fast memory. Notice that memory operations can recur multiple times depending on the loop in which they reside.
For instance, for two-level \strassen{},
\fromto{$T_{m}^{C_{\times}}=2\tau_{b}\lceil\frac{k/4}{k_c}\rceil \frac{m}{4}\frac{n}{4}$}
{$T_{m}^{C_{\times}}=2\lceil\frac{k/4}{k_c}\rceil \frac{m}{4}\frac{n}{4}\tau_{b}$}
denotes the cost of reading and writing the $\frac{m}{4}\times \frac{n}{4}$ submatrices of $C$ as intermediate result inside the micro-kernel. This is a step function proportional to $k$, because submatrices of $C$ are used to accumulate the \rankk{} update in the 5th loop in \figref{fig:side_by_side}(right).

%

\subsection{Discussion}
From the analysis summarized in~\figref{tab:breakdown} we can 
make predications about the relative performance of the various implementations.  It helps to also view the predictions as graphs, 
which we give in~\figref{fig:model}, using parameters that capture the architecture described in Section~\ref{sec:perf:single_node}.

\begin{itemize}
	\item 
	Asymptotically, the two-level \strassen\ implementations outperform corresponding one-level \strassen\ implementations, which in turn outperform the traditional \dgemm\ implementation.
	\NoShow{\item
	Asymptotically, when $ m $, $ n $, and $ k $ are all large, \XXXstrassen\ may actually outperform the other implementations. 
	This is because the use of temporary buffers 
	However, this is for matrices large enough that we do not see this in~\figref{fig:model}.
    }
	\item
	The graph for \mbox{$ m = k = n $, 1 core}, shows that for smaller square matrices, \ABCstrassen\ outperforms \ABXstrassen, but for larger square matrices this trend reverses.  This holds for both one-level and two-level \strassen{}.   The reason is that for small $ k $ \ABCstrassen\ reduced the number of times the temporary matrix $ M $ needs to be brought in from memory to be added to submatrices of $ C $.  For large $ k $, it increases the number of times the elements of those submatrices of $ C $ themselves are moved in and out of memory.
	\item
	The graph for \mbox{$ m = n = 16000$, $k$ varies, 1 core,} is particularly interesting: it shows that for $ k $ equal to the appropriate multiple of $ k_C $ ($k=2k_C$ for one-level and $ k=4k_C$ for two-level) \ABCstrassen\ performs dramatically better than the other implementations, as expected.
\end{itemize}
\NoShow{7. (Wrong) Conclusion:
\begin{itemize}
\item 0--2$k_c$: \dgemm\
\item 2$k_c$--$const\times n_c$: \ABCstrassen{} or \ABXstrassen{}
\item $n>const\times n_c$: \XXXstrassen{}
\end{itemize}
}%
The bottom line: depending on the problem size, a different implementation 
may have its advantages.

\NoShow{
The performance model help us understand of why the different implementations perform as observed. It establishes that we understand what is going on as the data is being moved within the fast and slow memory. 
}





%% file: 05performance.tex
\input setup

%% file: setup.tex
\begin{figure*}[htp!]
	\centering
	\includegraphics[width=0.43\textwidth]{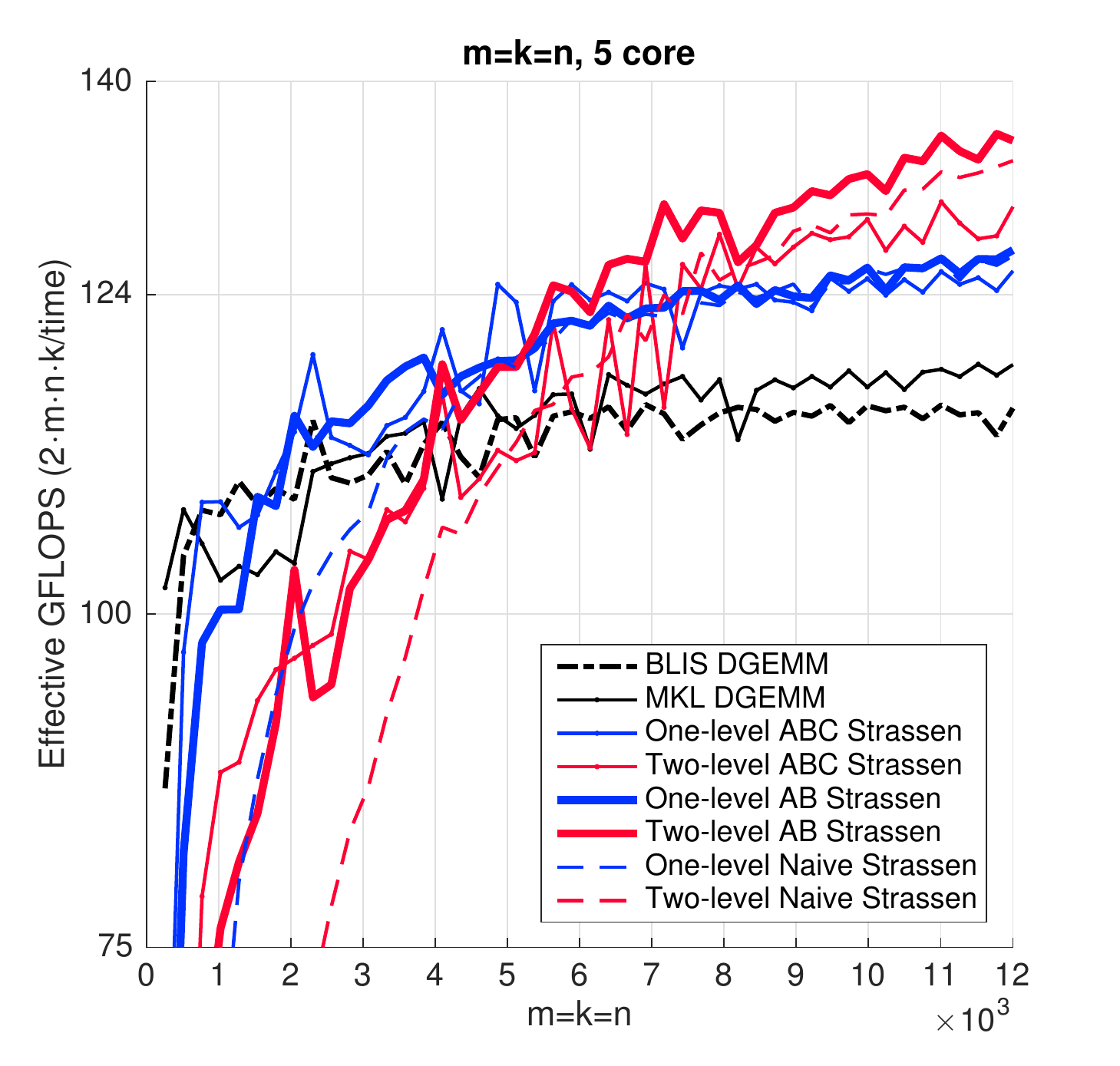}
	\includegraphics[width=0.43\textwidth]{figures/outsquare_10core.pdf}
	\includegraphics[width=0.43\textwidth]{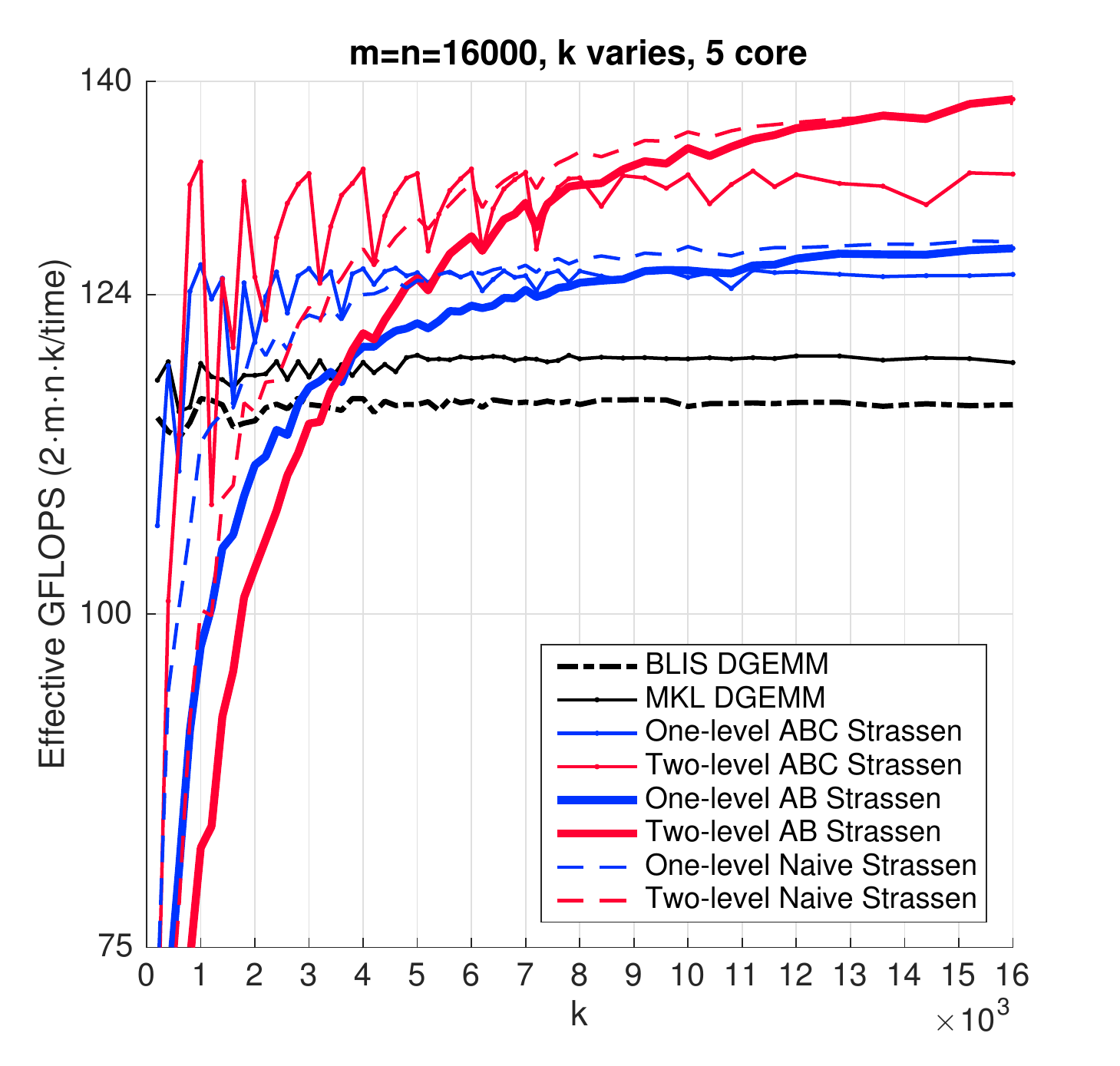}
	\includegraphics[width=0.43\textwidth]{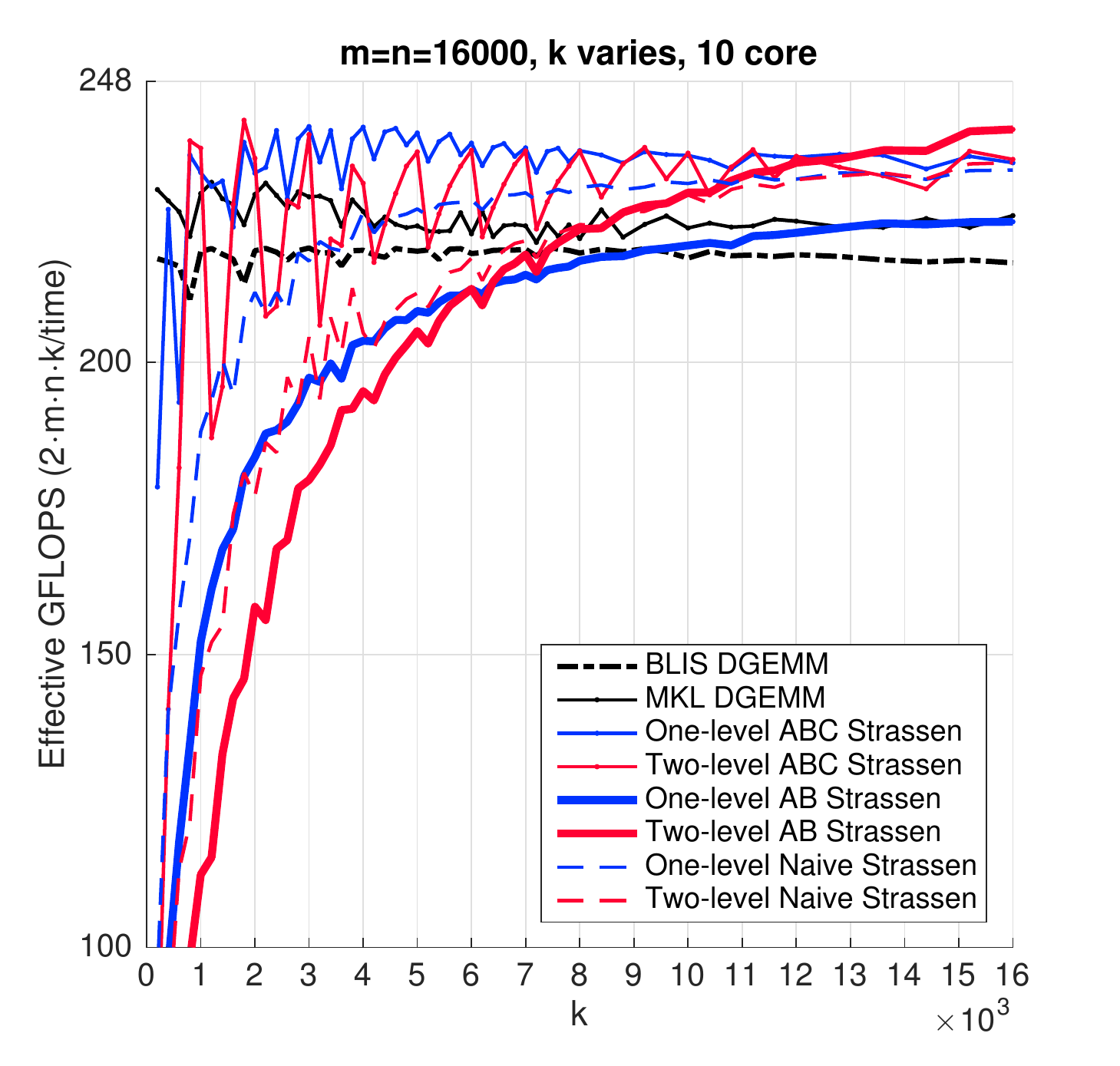}
	\includegraphics[width=0.43\textwidth]{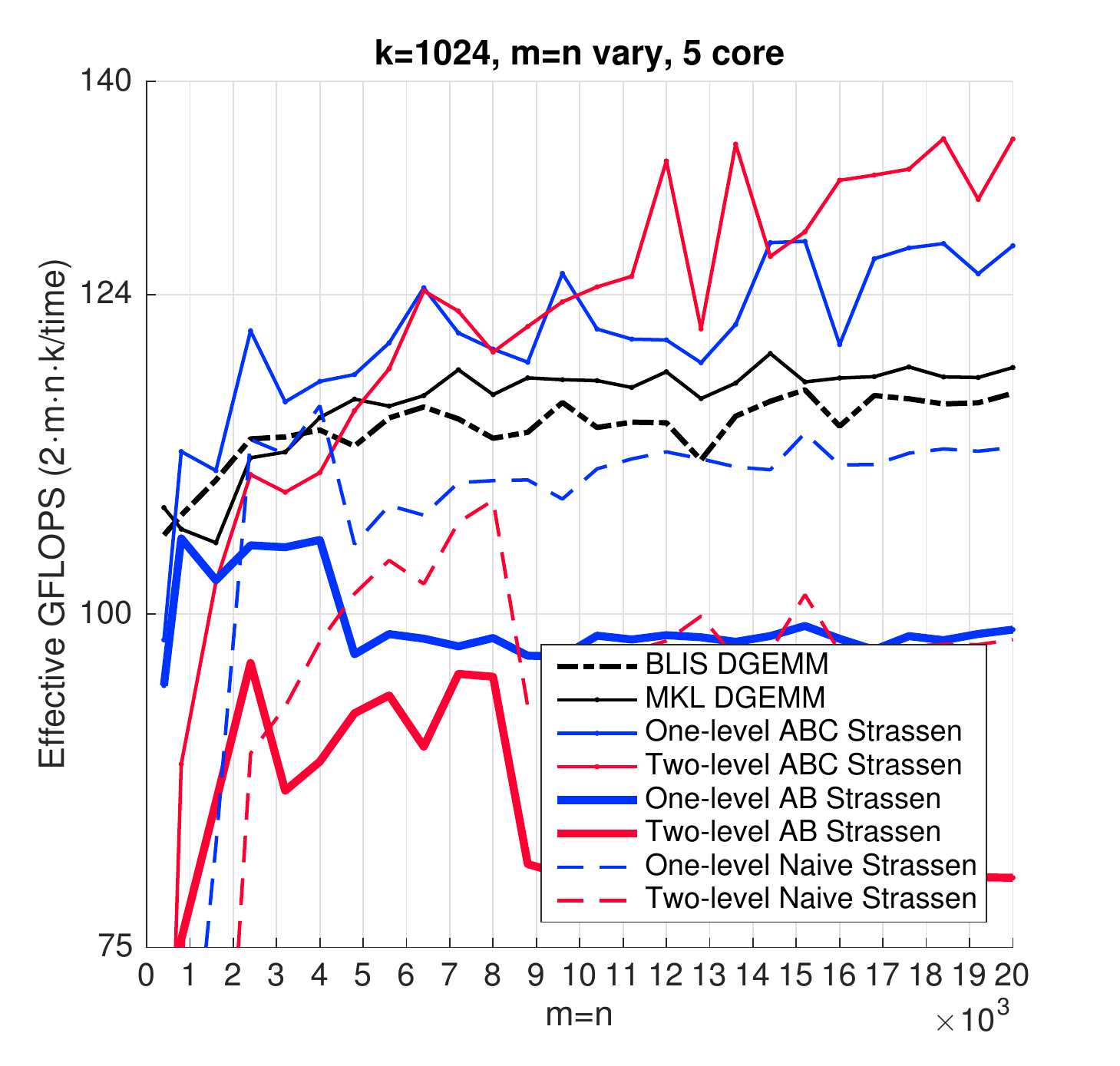}
	\includegraphics[width=0.43\textwidth]{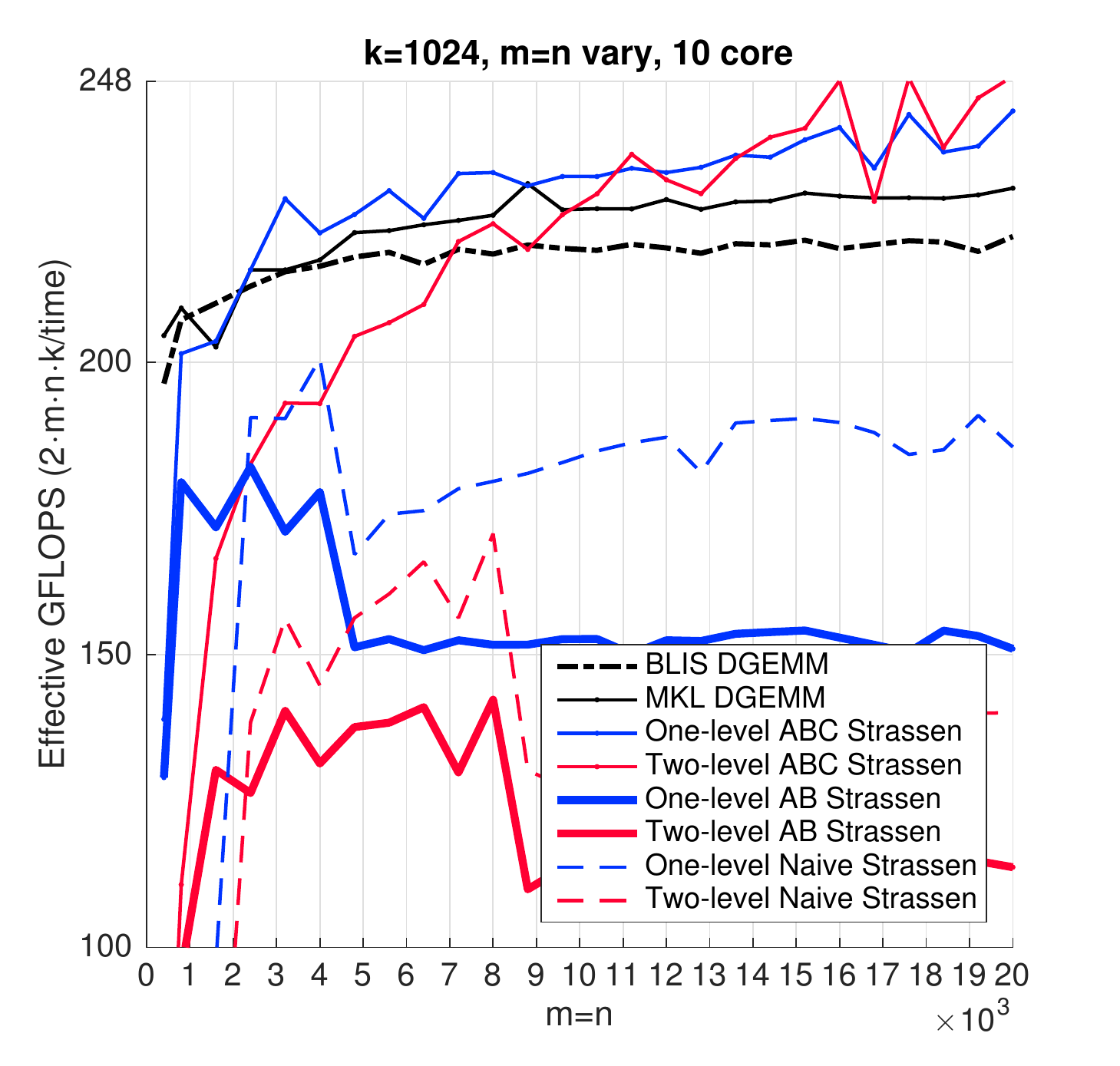}
	\caption{Performance of the various implementations on an Intel\textregistered\ Xeon\textregistered\ E5 2680 v2 (Ivybridge) processor.  
	}
	\label{fig:single_node}
\end{figure*} 

We give details on the performance experiments for our implementations.
The current version of \strassen\ \dgemm\ is designed for the Intel\textregistered\ Xeon\textregistered\ (Sandy-Bridge/Ivy-Bridge) processors and
Intel\textregistered\ Xeon Phi\texttrademark\ coprocessor (MIC architecture, KNC).  In addition, we incorporate our implementations in a distributed memory \gemm{}.

\subsection{Single node experiments}
\label{sec:perf:single_node}

\subsubsection*{Implementation}

The implementations are in {\tt C}, utilizing {\tt SSE2} and {\tt AVX} intrinsics and assembly, compiled with the Intel\textregistered\ {\tt C++} Compiler version 15.0.3 with optimization flag {\tt -O3}. 
In addition, we compare against the standard \BLIS\  implementation (Version 0.1.8) from which our implementations are derived as well as Intel\textregistered\ Math Kernel Library (Intel\textregistered\ MKL) \dgemm\ (Release Version 11.2.3) \cite{IntelMKL}.

\subsubsection*{Target architecture}

We measure the CPU performance results on the Maverick system at the Texas Advanced Computing Center (TACC).  Each node of that system consists of a  
dual-socket Intel\textregistered\ Xeon\textregistered\ E5-2680 v2 (Ivy Bridge) processors with 12.8GB/core of memory
(peak Bandwidth: 59.7 GB/s with four channels) and a three-level cache: 32KB L1 data cache, 256KB L2 cache and 25.6MB L3 cache.
The stable CPU clockrate is 3.54GHz when a single core is utilized (28.32 GFLOPS peak, marked in the graphs) and 3.10GHz when five or more cores are in use (24.8 GLOPS/core peak). 
To set thread affinity and to ensure the computation and the memory allocation all reside on 
the same socket, we use {\tt KMP\_AFFINITY=compact}.

We choose the parameters $n_R=4$, $m_R=8$, $k_C=256$, $n_C=4096$ and $m_C=96$. This makes the size of the packing buffer
$ \widetilde A_i $ 192KB and 
$ \widetilde B_p $ 8192KB, which then fit the L2 cache and L3 cache, respectively.  These parameters are consistent with parameters used for the standard \BLIS\ \dgemm\ implementation for this architecture.

Each socket consists of 10 cores, allowing us to also perform multi-threaded experiments.  Parallelization 
is implemented mirroring that described in~\cite{BLIS3}, using OpenMP directives that parallelize the $3^{rd}$ loop around the micro-kernel in Figure~\ref{fig:side_by_side}.  

\subsubsection*{Results}

\NoShow{
	\begin{figure*}[htp]
	\centering
	\includegraphics[width=0.32\textwidth]{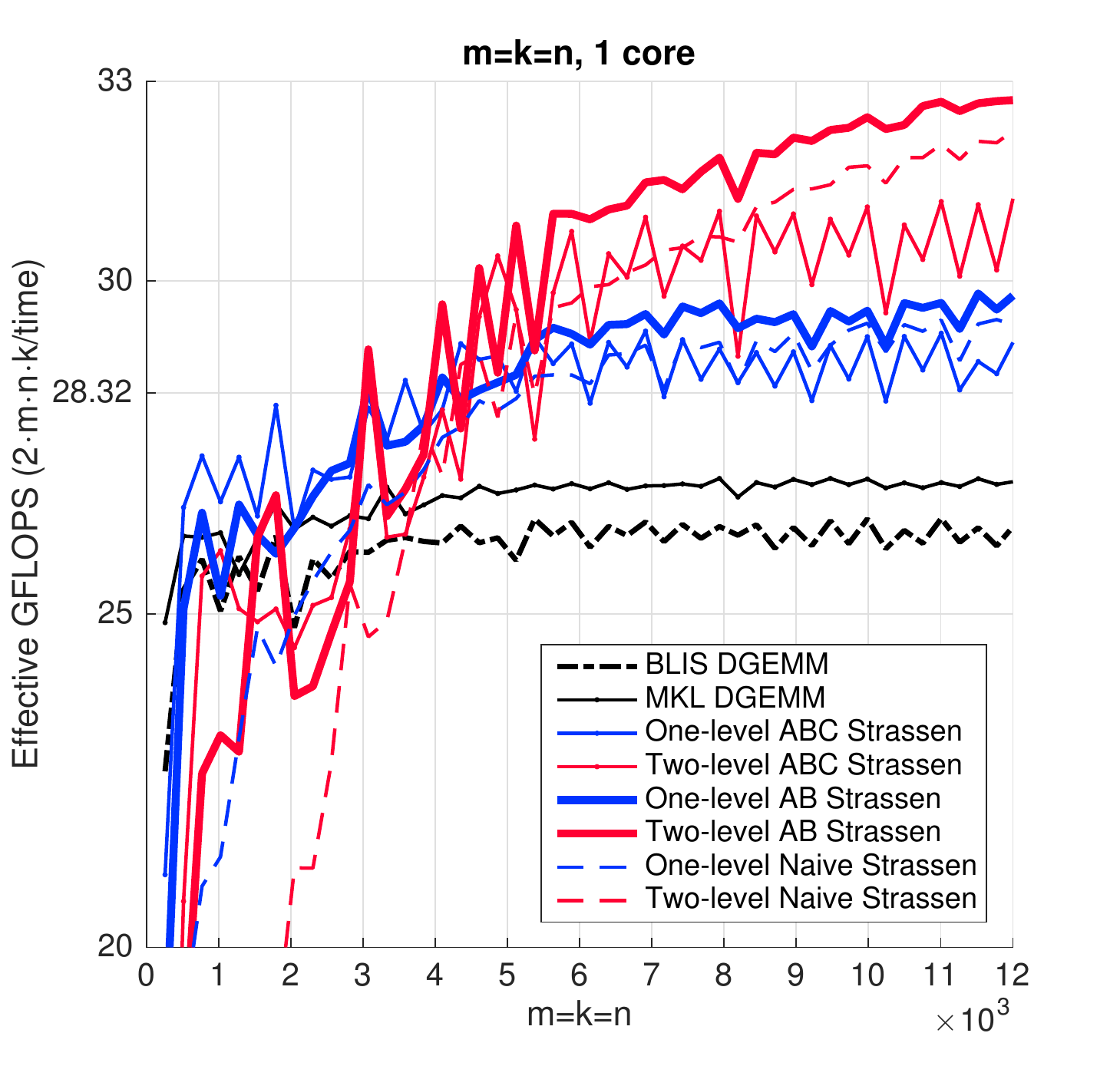}
	\includegraphics[width=0.32\textwidth]{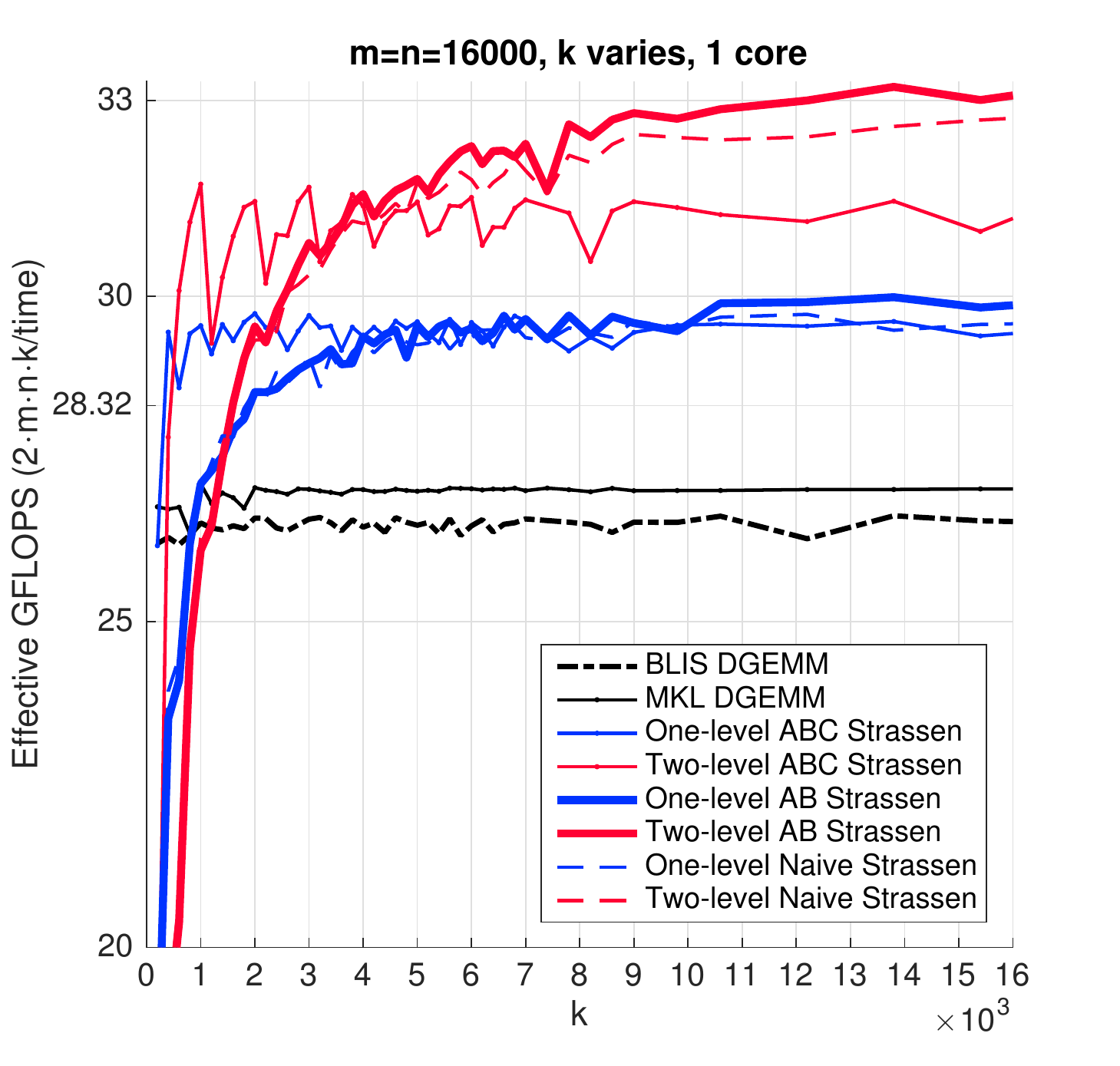}
	\includegraphics[width=0.32\textwidth]{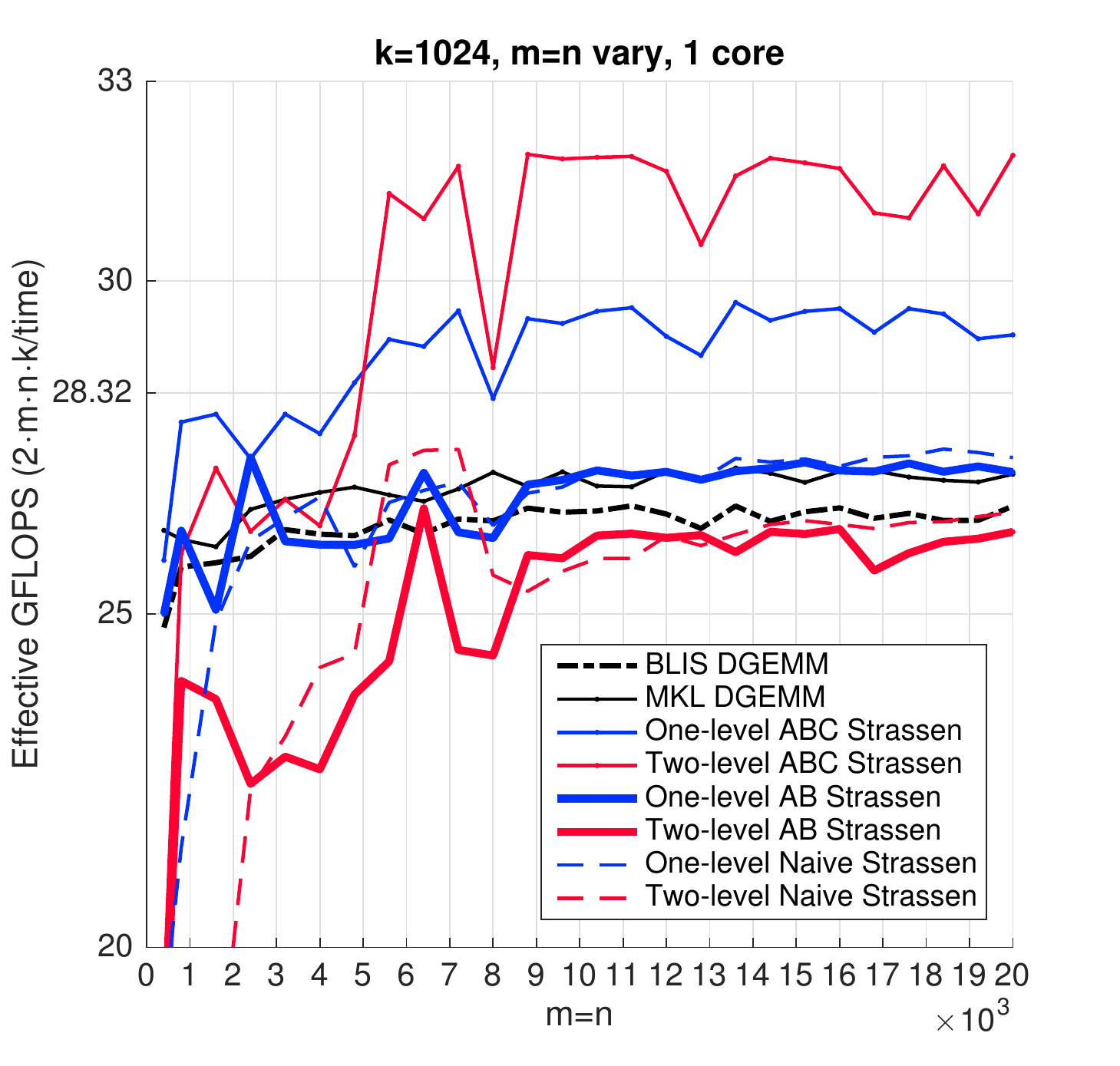}
	\\
	\includegraphics[width=0.32\textwidth]{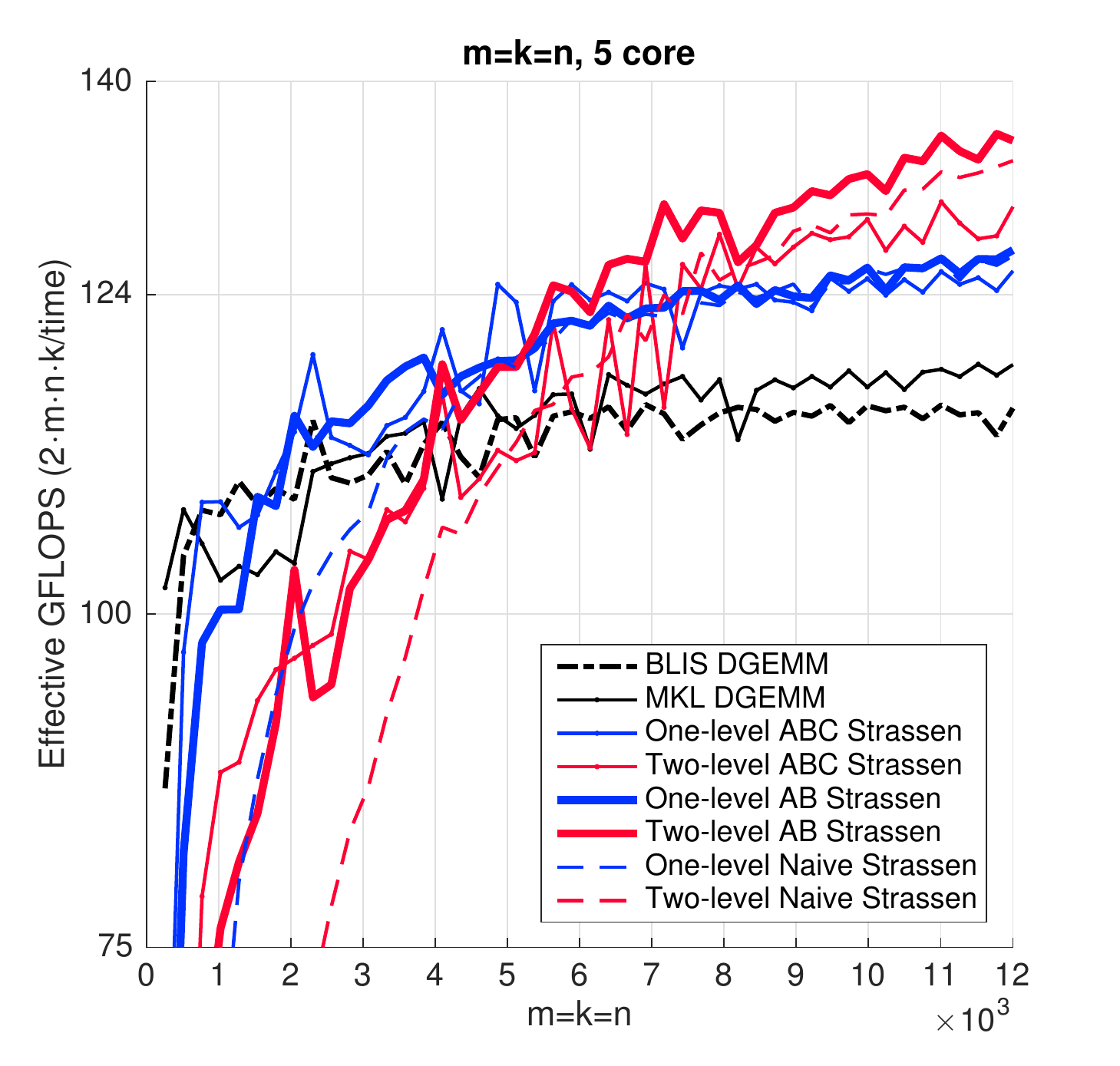}
	\includegraphics[width=0.32\textwidth]{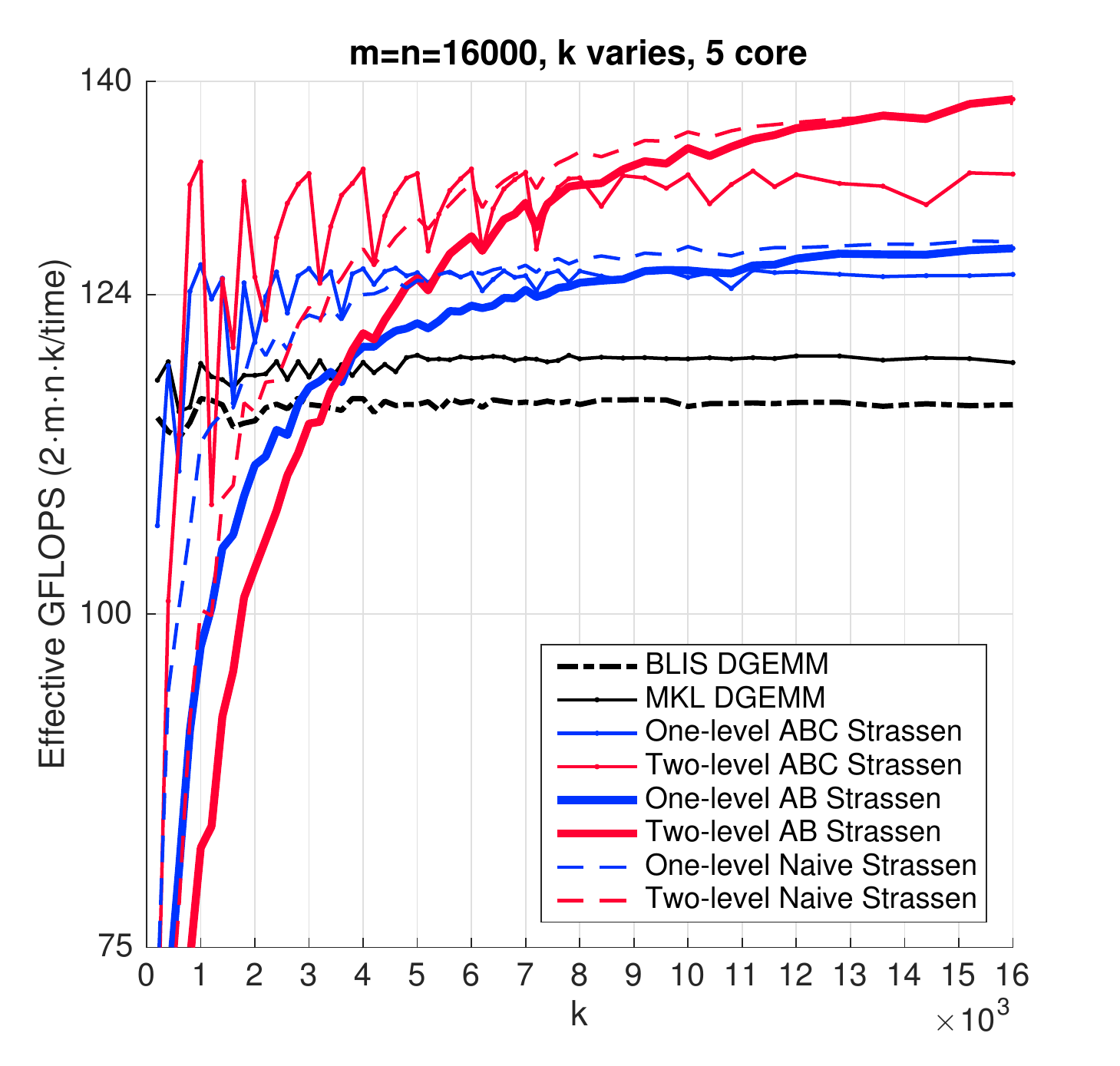}
	\includegraphics[width=0.32\textwidth]{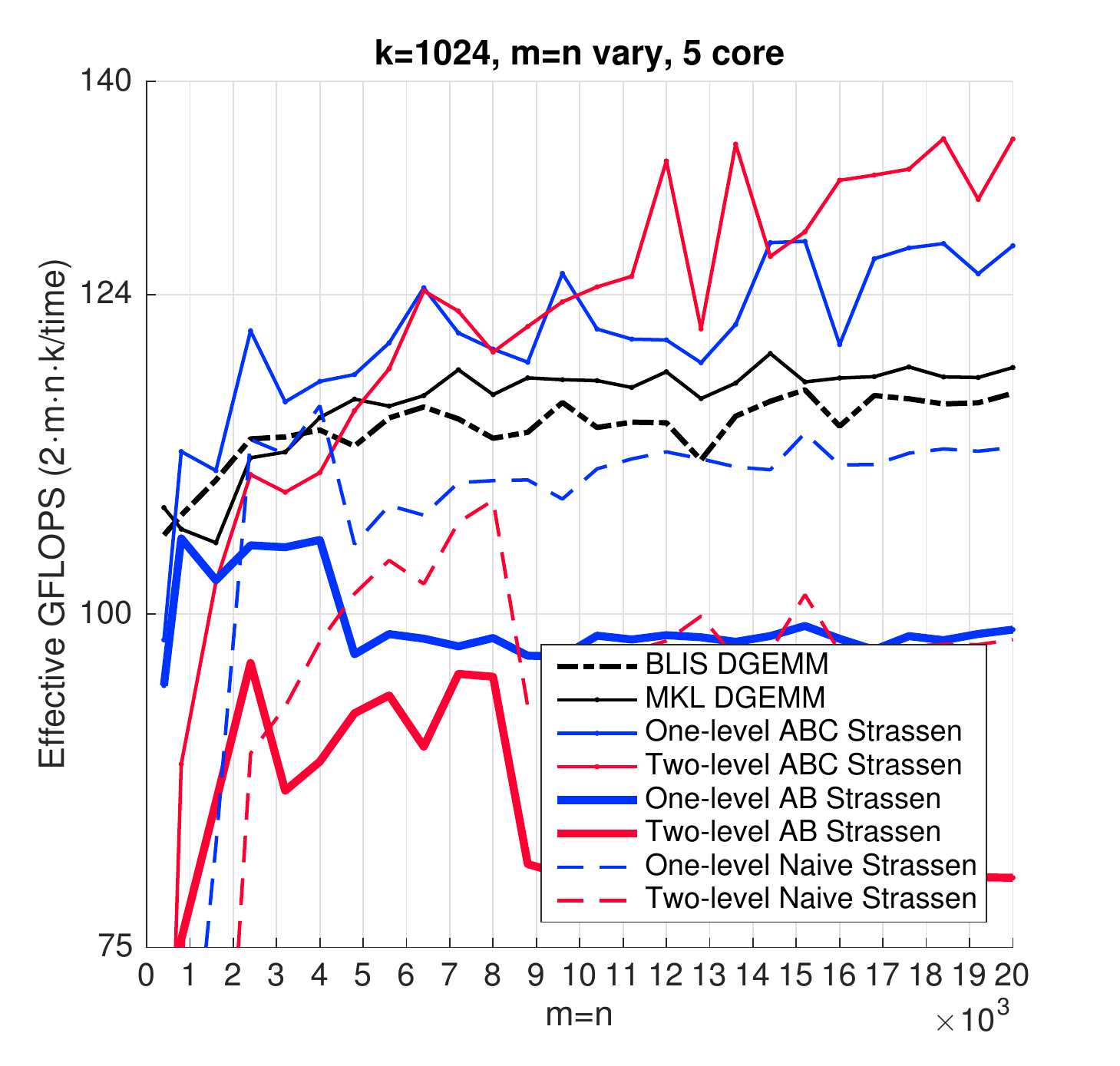}
	\\
	\includegraphics[width=0.32\textwidth]{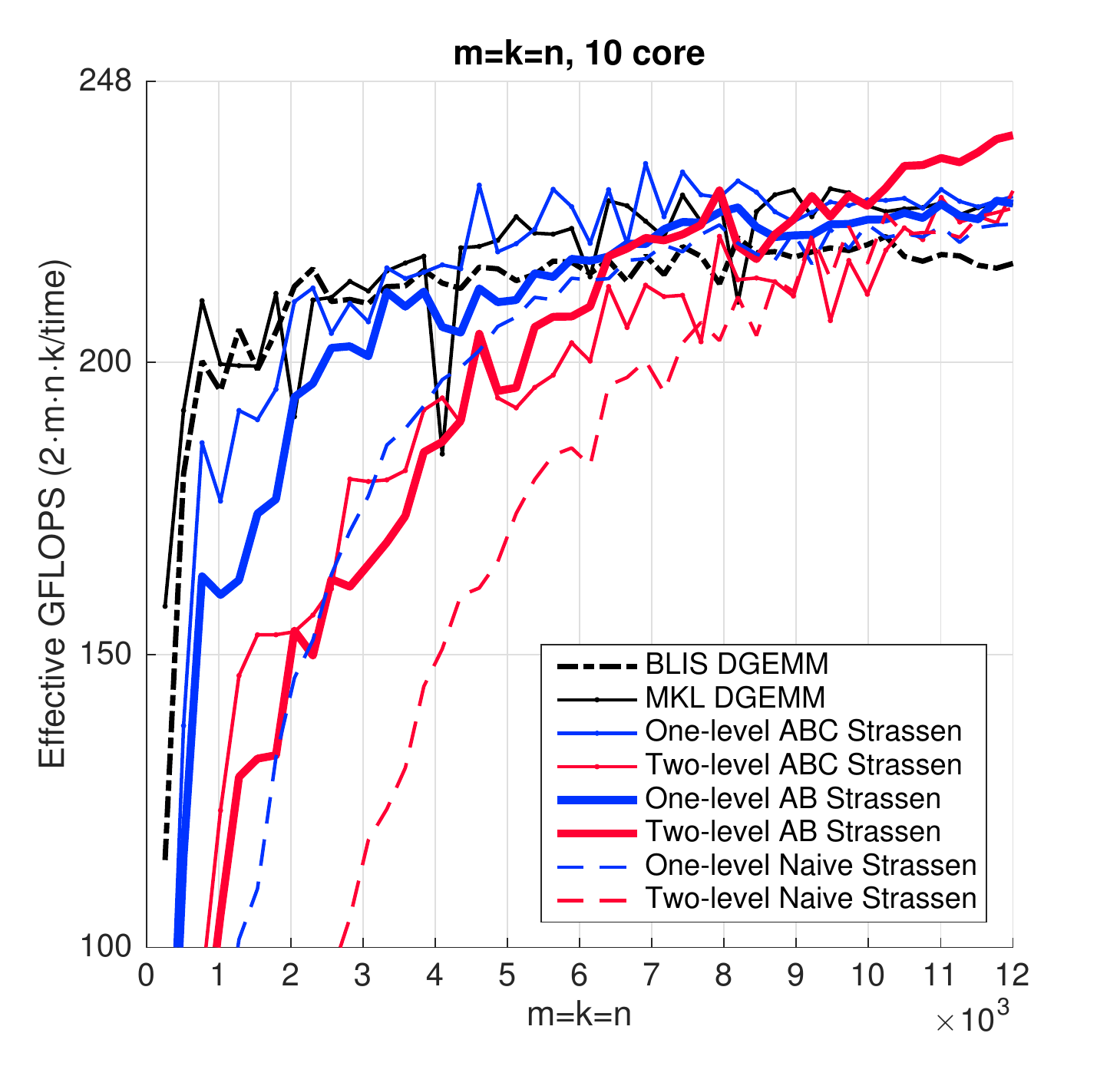}
	\includegraphics[width=0.32\textwidth]{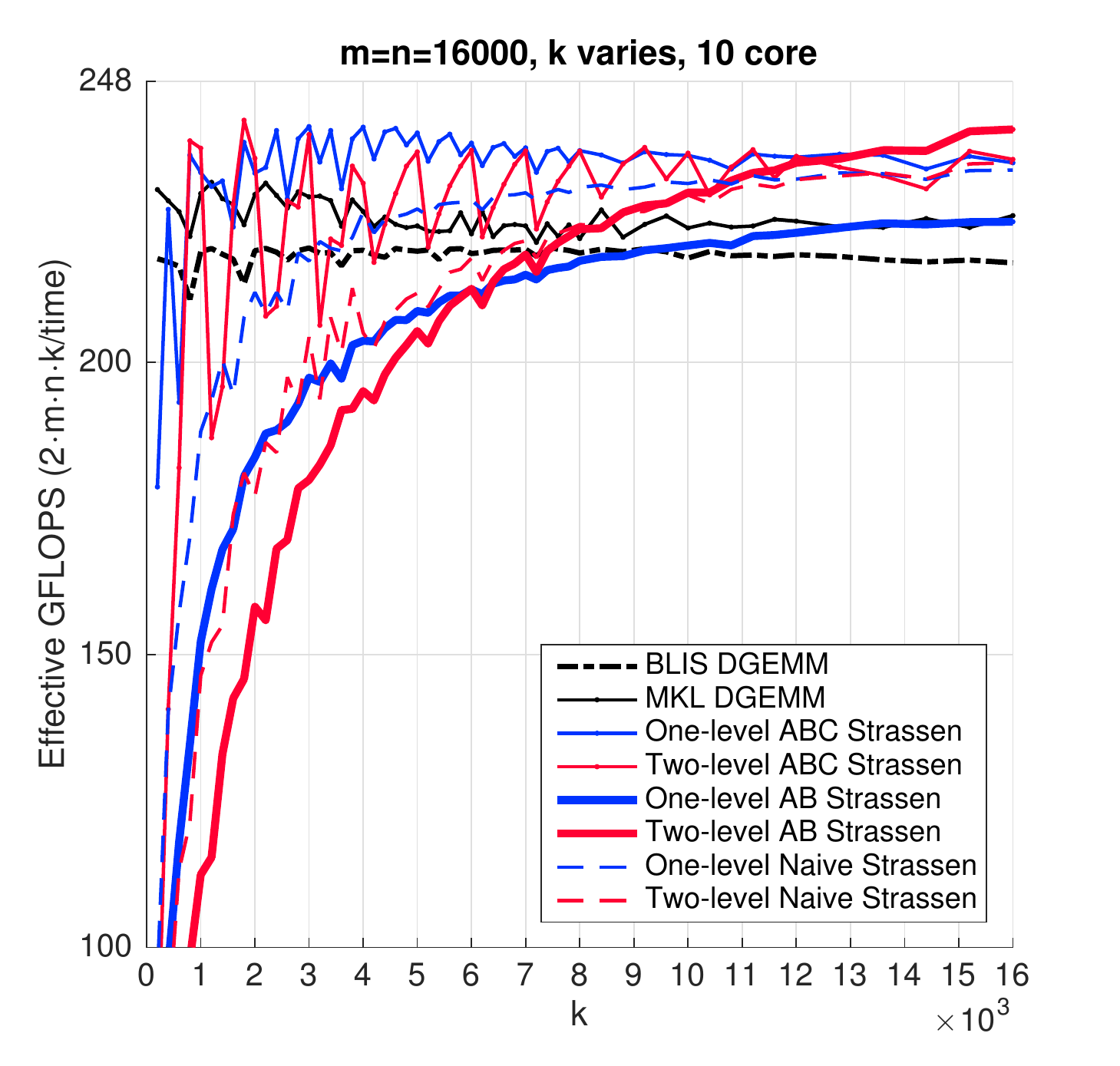}
	\includegraphics[width=0.32\textwidth]{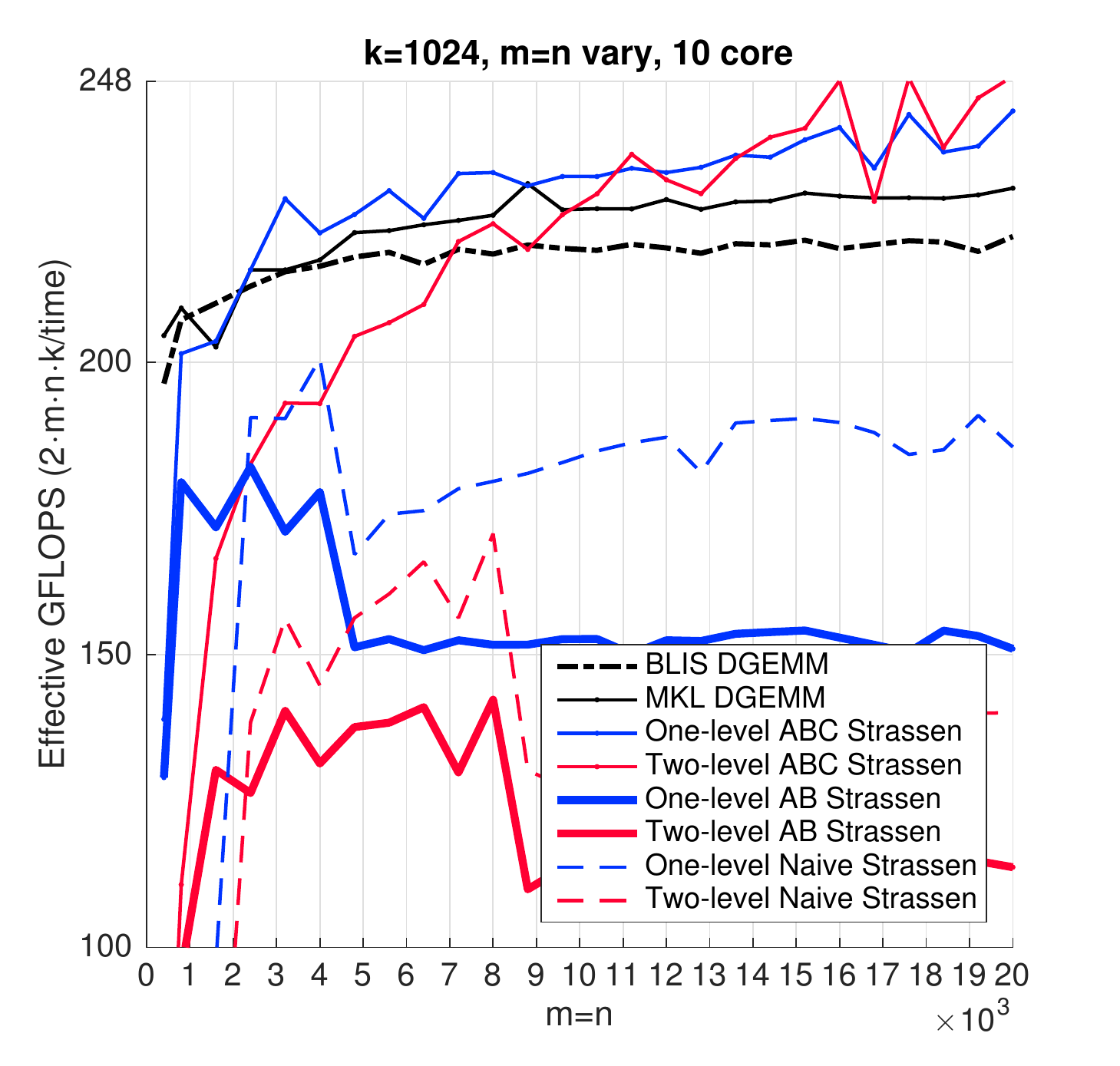}
	\caption{Performance of the various implementations on an Intel Xeon E5 2680 v2 (Ivybridge) processor.  Notice that the range of the y-axis does not start at zero to make the graphs more readable.  For single core experiments, 28.32 marks theoretical peak performance.
	}
	\label{fig:single_node}
\end{figure*} 
}

Results when using single core are presented in \figref{fig:model} (right column).
As expected, eventually two-level \ABXstrassen\ performs best, handily beating conventional \dgemm.  
The exception is the case where $ k $ is fixed to equal $ 1024 = 4 \times k_C $, which is the natural blocking size for a two-level \strassen\ based on our ideas.  For those experiments \ABCstrassen\ wins out, as expected.
These experiments help validate our model.

\figref{fig:single_node} reports results for five and ten cores, all within the same socket.   We do not report results for twenty cores (two sockets), since this results in a substantial performance reduction for all our implementations, including the standard \BLIS\ \dgemm, relative to  Intel\textregistered\ MKL \dgemm.  This exposes a performance bug in \BLIS\ that has been reported.

When using many cores, memory bandwidth contention affects the performance of the various \strassen\ implementations, reducing the benefits relative to a standard \dgemm\ implementation.

\subsection{Many-core experiments}
To examine whether the techniques scale to a large number of cores, we port on implementation of one-level \ABCstrassen\ to the Intel\textregistered\ Xeon Phi\texttrademark coprocessor.

\subsubsection*{Implementation}

The implementations of \ABCstrassen{} are in {\tt C} and {\tt AVX512} intrinsics and assembly, compiled with the Intel {\tt C} compiler version 15.0.2 with optimization flag {\tt -mmic -O3}.  The \BLIS\ and \ABCstrassen\ both parallelize the $2^{nd}$ and $3^{rd}$ loop around the micro-kernel, as described for \BLIS\ in~\cite{BLIS3}.

\subsubsection*{Target architecture}

We run the Intel\textregistered\ Xeon Phi\texttrademark\ coprocessor performance experiments on the SE10P Coprocessor incorporated into nodes of the Stampede system at TACC. This coprocessor
has a peak performance of 1056GFLOPS (for 60 cores/240 threads used by BLIS) and 8GB of GDDR5 DRAM with a peak bandwidth of 352GB/s. It has 512KB L2 cache, but no L3 cache.

We choose the parameters $n_R=8$, $m_R=30$, $k_C=240$, $n_C=14400$ and $m_C=120$. This makes the size of the packing buffer
$ \widetilde A_i $ 225KB and 
$ \widetilde B_p $ 27000KB, which fits L2 cache and main memory separately (no L3 cache on the Intel\textregistered\ Xeon Phi\texttrademark\ coprocessor).  These choices are consistent with those used by \BLIS\ for this architecture.

\subsubsection*{Results}

\begin{figure}[!tp]
	\centering
	\includegraphics[width=0.43\textwidth]{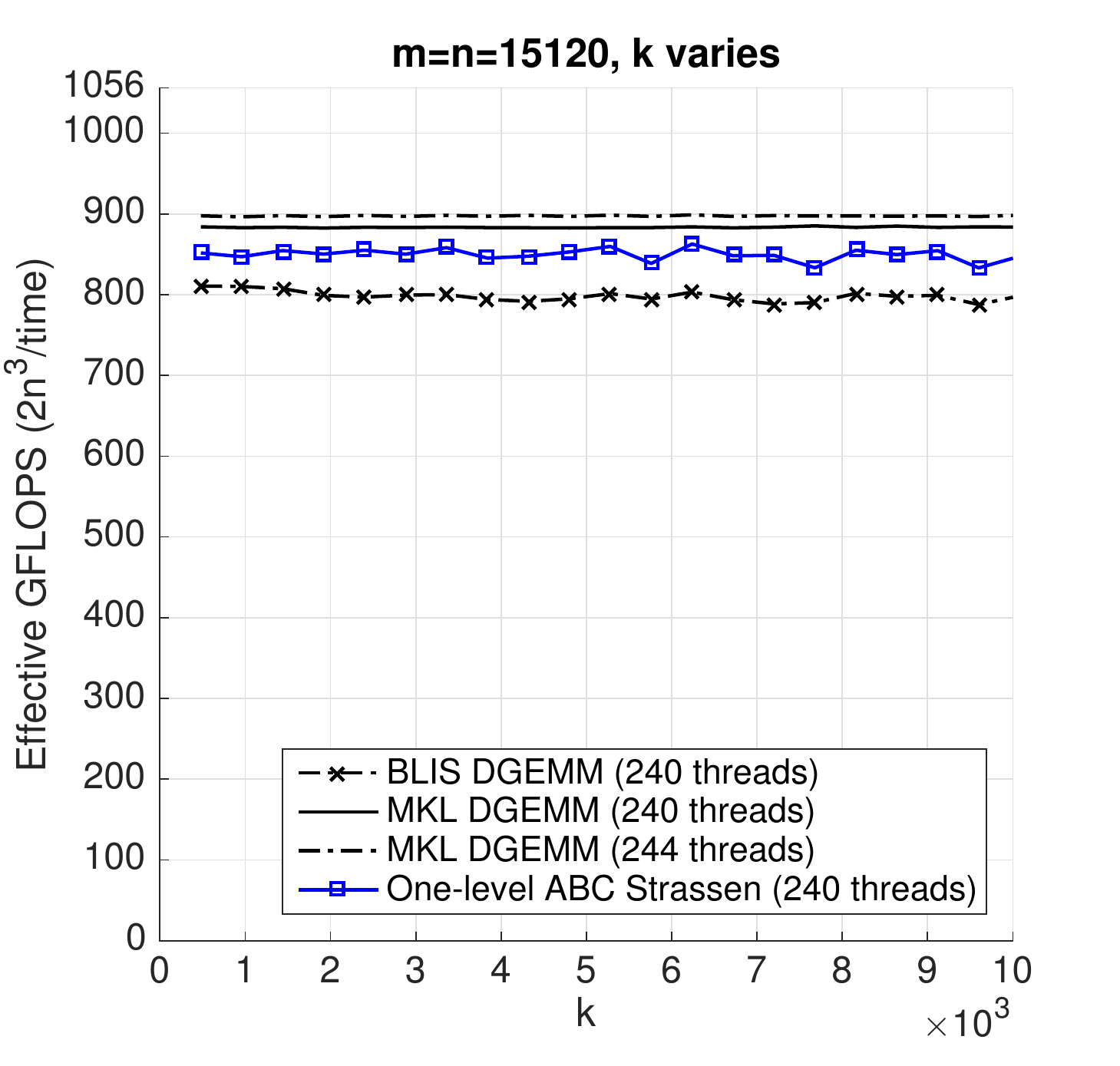}
	\caption{Performance of one-level \ABCstrassen, BLIS, and Intel\textregistered\ MKL, on an Intel\textregistered\ Xeon Phi\texttrademark\ coprocessor (up to 61 cores with 244 threads).
	}
	\label{fig:mic}
\end{figure}

As illustrated in \figref{fig:mic}, relative to the \BLIS\ \dgemm\ implementation, the one-level \ABCstrassen\ shows a nontrivial improvement for a \rankk{} update with a fixed (large) matrix $ C $.  While the \BLIS\ implementation on which our implementation of \ABCstrassen\ used to be highly competitive with Intel\textregistered\ MKL's \dgemm\ (as reported in~\cite{BLIS3}), 
recent improvements in that library demonstrate that the \BLIS\ implementation needs an update.  We do not think there is a fundamental reason why our observations cannot be used to similarly accelerate Intel\textregistered\ MKL's \dgemm.


\subsection{Distributed memory experiments}

Finally, we demonstrate how the \ABCstrassen\ implementation can be used to accelerate a distributed memory implementation of \dgemm.
\subsubsection*{Implementation}

We implement the Scalable Universal Matrix Multiplication Algorithm (SUMMA)~\cite{SUMMA} with MPI.  This algorithm distributes the algorithm to a mesh of MPI processes using a 2D block cyclic distribution.  The multiplication
is broken down into a sequence of \rankb{} updates, %
{
\begin{eqnarray*}
C &:=& A B + C = \left(\begin{array}{c | c | c }
A_0 & \cdots & A_{K-1}
\end{array}
\right)
\left(
\begin{array}{c}
B_0 \\ \hline
\vdots \\ \hline
B_{K-1}
\end{array}
\right) + C \\
&=&
A_0 B_0 + \cdots + A_{K-1} B_{K-1} + C
\end{eqnarray*}%
}%
where each $ A_p $ consists of (approximately) $ b $ columns and each $ B_p $ consists of (approximately) $ b $ rows.
For each \rankb{} update  $ A_p $ is  broadcast within rows of the mesh and  $ B_p $ is broadcast within columns of the mesh, after which locally a \rankb{} update with the arriving submatrices is performed to update the local block of $ C $.  


\subsubsection*{Target architecture}

The distributed memory experiments are performed on the same machine described in Section~\ref{sec:perf:single_node}, using the 
{\tt mvapich2} version 2.1~\cite{MVAPICH2} implementation of MPI.
Each node has two sockets, and each socket has ten cores.  



\begin{figure}[!tp]
	\centering
	\includegraphics[width=0.43\textwidth]{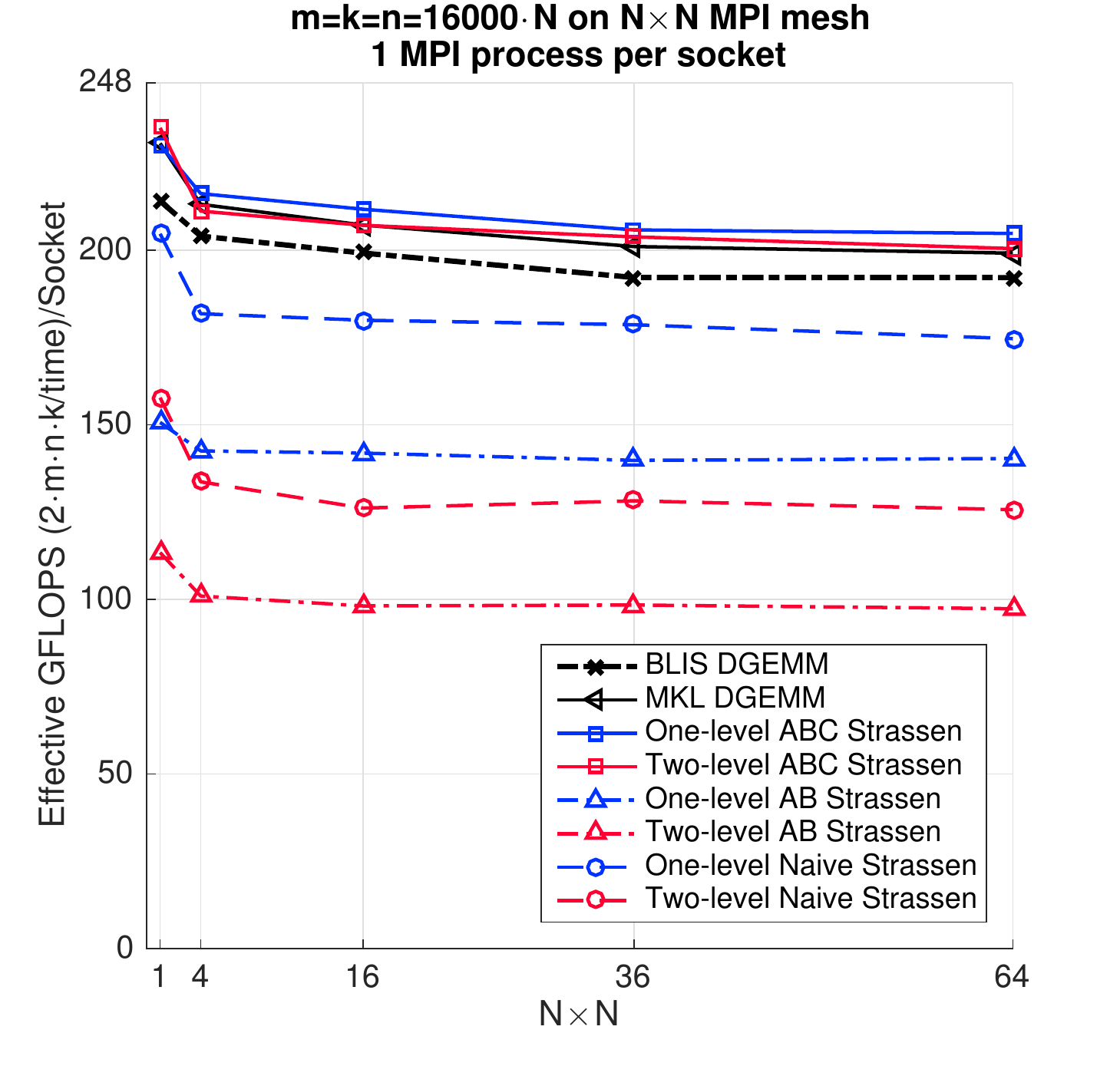}
	\caption{Performance of the various implementations on distributed memory (weak scalability).
	}
	\label{fig:summa}
\end{figure} 

\subsubsection*{Results}
\figref{fig:summa} reports weak scalability \fromto{}{on up to 32 nodes (64 sockets, 640 cores)}.
For these experiments we choose the MPI mesh of processes to be square, with one MPI process per socket, and attained thread parallelism among the ten cores in a socket within BLIS, Intel\textregistered\ MKL, or any of our \strassen\ implementations.

It is well-known that the SUMMA algorithm is weakly scalable in the sense that efficiency essentially remains constant if the local memory dedicated to matrices $ A $, $ B $, $ C $, and temporary buffers is kept constant. 
For this reason, the local problem size is fixed to equal $m=k=n=16000$ so that the global problem becomes $ m=k=n=16000\times N $ when an $ N \times N $ mesh of sockets (MPI processes) is utilized. 
As expected, the graph shows that the SUMMA algorithm is weakly scalable regardless of which local \gemm\ algorithm is used.
The local computation within the SUMMA algorithm matches the shape for which \ABCstrassen\ is a natural choice when the \rankk{} updates are performed with $ b = 1024 $.  For this reason, the one-level and two-level \ABCstrassen\ implementations achieve the best performance.

What this experiment shows is that the benefit of using our \strassen{} implementations can be easily transferred to other algorithms that are rich in large \rankk{} updates.

%% file: 06conclusion.tex
\NoShow{
	\input related

First try to break the BLAS interface to embrace Stassen algorithm

Build a thorough performance model

The readers may want to compare our single core performance with other relate work.

multi-core performance, totally orthogonal...
}

We have presented novel insights into the implementations of \strassen\ that greatly reduce overhead that was inherent in previous formulations and had been assumed to be insurmountable.   These insights have yielded a family of algorithms that outperform conventional high-performance implementations of \gemm\ as well as naive implementations.  We develop a model that predicts the run time of the various implementations.  Components that are part of the BLIS framework for implementing BLAS-like libraries are modified to facilitate implementation.  Implementations and performance experiments are presented that verify the performance model and demonstrate performance benefits for single-core, multi-core, many-core, and distributed memory parallel implementations.  Together, this advances more than 45 years of research into the theory and practice of Strassen-like algorithms.

{\color{black} Our analysis shows that the \ABCstrassen\ implementation fulfills our claim that \strassen\ can outperform classical \gemm\ for small matrices and small $ k $ while requiring no temporary buffers beyond those already internal to high-performance \gemm\ implementations.  The \ABXstrassen\ algorithm becomes competitive once $ k $ is larger.  It only requires a $ \frac{m}{2^L} \times \frac{n}{2^L} $ temporary matrix for an $ L $-level \strassen\ algorithm.}

A number of avenues for further research and development naturally present themselves.  
\begin{itemize}
	\item 
	The GotoBLAS approach for \gemm\ is also the basis for high-performance implementations of all level-3 BLAS~\cite{1377607} and the BLIS framework has been used to implement these with the same micro-kernel and modifications of the packing routines that support \dgemm.  
	This presents the possibility of creating Strassen-like algorithms for some or all level-3 BLAS.
	\item
	Only the \ABCstrassen\ algorithm has been implemented for the Intel\textregistered\ Xeon Phi\texttrademark\ coprocessor.  While this demonstrates that parallelism on many-core architectures can be effectively exploited, a more complete study needs to be pursued.  Also, the performance improvements in Intel\textregistered\ MKL for that architecture need to be duplicated in BLIS and/or the techniques incorporated into the Intel\textregistered\ MKL library.
	\item
	Most of the blocked algorithms that underlie LAPACK and ScaLAPACK~\cite{ScaLAPACK} cast computation in terms of \rankk{} updates.  It needs to be investigated how the \ABCstrassen\ algorithm can be used to accelerate these libraries.
	\item
	Distributed memory implementations of Strassen's algorithms have been proposed that incorporate several levels of Strassen before calling a parallel SUMMA or other distributed memory parallel \gemm\ implementation~\cite{PPL2}.  On the one hand, the performance of our approach that incorporates \strassen\ in the local \gemm\ needs to be compared to these implementations.  On the other hand, it may be possible to add a local \strassen\ \gemm\ into these parallel implementations.  Alternatively, the required packing may be incorporated into the communication of the data.
	\item
	A number of recent papers have proposed multi-threaded parallel implementations that compute multiple submatrices $ M_i $ in parallel~\cite{D'alberto:2011:EPM:2049662.2049664}.  Even more recently, new practical Strassen-like algorithms have been proposed together with their multi-threaded implementations~\cite{StrassenBenson}.  How our techniques compare to these and whether they can be combined needs to be pursued.  It may also be possible to use our cost model to help better schedule this kind of task parallelism.
\end{itemize}
These represent only a small sample of new possible directions.

%% file: related.tex
\subsection{Related Work}
\subsubsection*{Numerical stability}

\cite{DemmelStrassenStable2007,BallardStrassenStable15} paper.

\subsubsection*{Implementation}
\cite{StrassenBenson,D'Alberto:2009:AWM:1486525.1486528} paper.

Benson: no performance benefit for small problem size...

just describe OR performance comparison?

\subsubsection*{Future work?}
1.Strassen like (2x2x3, etc)
2.Other architectures
3.task-level parallelism

Note that \Strassen{} is limited by communication lower bounds \cite{Ballard2013}, and we attain them by our practical approach.